\documentclass{article}
\usepackage{arxiv}
\usepackage[utf8]{inputenc} 
\usepackage[T1]{fontenc}    
\usepackage{hyperref}       
\usepackage{url}            
\usepackage{booktabs}       
\usepackage{amsfonts}       
\usepackage{nicefrac}       
\usepackage{microtype}      
\usepackage{lipsum}		
\usepackage{graphicx}
\usepackage{natbib}
\usepackage{doi}
\usepackage{array}
\usepackage{dcolumn}
\usepackage{bm}
\usepackage{subcaption}
\usepackage[utf8]{inputenc}
\usepackage[T1]{fontenc}
\usepackage{enumitem}
\usepackage{booktabs}
\usepackage{gensymb}
\usepackage{amsmath}
\usepackage{mathptmx} 
\usepackage{caption}
\title{Improvement of NACA6309 Airfoil with Passive Air-Flow Control by using Trailing Edge Flap}

\author{\href{https://orcid.org/0000-0000-0000-0000}{\includegraphics[scale=0.06]{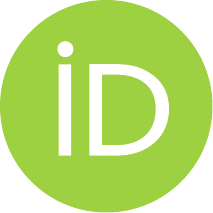}\hspace{1mm}Mahadi Hasan Shanto} \\
	Department of Mechanical Engineering\\
	Shahjalal University of Science and Technology\\
	Sylhet-3114, Bangladesh\\
	\texttt{mahadishanto27@gmail.com} \\
 	\And
   \href{https://orcid.org/0009-0002-4907-031X}{\includegraphics[scale=0.06]{orcid.pdf}\hspace{1mm}Sayed Tanvir Ahmed}\thanks{Corresponding author: sytanvir.mech@gmail.com} \\
	Department of Mechanical Engineering\\
	Shahjalal University of Science and Technology\\
	Sylhet-3114, Bangladesh\\
	\texttt{sytanvir.mech@gmail.com} \\
        \And
	\href{https://orcid.org/0000-0000-0000-0000}{\includegraphics[scale=0.06]{orcid.pdf}\hspace{1mm}A K M Ashikuzzaman} \\
	Department of Mechanical Engineering\\
	Shahjalal University of Science and Technology\\
	Sylhet-3114, Bangladesh\\
	\texttt{akmashik-mee@sust.edu} \\
}

\date{}

\hypersetup{
pdftitle={A template for the arxiv style},
pdfsubject={q-bio.NC, q-bio.QM},
pdfauthor={David S.~Hippocampus, Elias D.~Striatum},
pdfkeywords={First keyword, Second keyword, More},
}

\begin{document}
\maketitle
\begin{abstract}
When fossil fuel supplies can no longer be replenished and hence fossil fuel power generation becomes outdated, wind energy will become a vital solution to the impending energy crisis. A horizontal-axis wind turbine is a widely used technology that is highly dependent on the design of high-performing airfoils. In this paper, we have studied the performance of the NACA6309 airfoil and designed it by modifying the airfoil with a trailing edge plain flap. Computational Fluid Dynamic (CFD) simulations are utilized for this purpose. We have designed sixteen configurations of NACA 6309 airfoil by using plain flaps at the trailing edge and studied their aerodynamic performance. After comparing the lift, drag, and lift-to-drag ratios, it is evident that the \(1^\circ\)  up-flap configuration generates the best output. In addition, the \(10^\circ\) down flap provides the worst performance among all configurations. Finally, pressure contours and velocity contours around the airfoils are presented, which describe the overall characteristics.
\end{abstract}

\keywords{Wind turbine airfoil \and trailing edge \and aerodynamic performance \and lift-drag ratio \and plain flap \and CFD}

\section{Introduction}
\setlength{\parindent}{0.2in}
\hspace{0.2in}
Wind technology is truly becoming the pinnacle of human advancement though the idea is arcane, the advancement is at large about not just the complexity of the technology but the impact on human life itself. With increasing research on wind technology, along with proper political decisions around the world, wind turbine technology, especially vertical axis is gaining more reliability and popularity as a viable power generation alternative. In this research, we will mainly focus on airfoil improvement used in three-bladed wind turbines for power generation purposes, also familiar as Horizontal Axis Wind Turbine (HAWT). The unorthodox counterpart of HAWT is VAWT which stands for Vertical Axis Wind Turbine. In the early 1970s, SANDIA Lab highlighted the necessary experimental results concerning HAWT and its aerodynamical performance, and structural integrity and analyzed the overall system, which was a 17-m turbine generator system \cite{osti_7356038}. Though the technology was in literature from that time, it has been recently rediscovered to be an interesting solution to cope with ever-increasing power demand; yet HAWT is still the dominating one.  Hence, to focus on the improvement of such a power generation system, flow control techniques appeared to be the most robust. Many interesting flow control techniques can be found in the literature. The passive pitch control technique results in an improvement in power generation lifetime, reduction in instabilities, and structural performance (Motley and Barber 2014). Secondly, by concentrating flow in the rotor zone which can be achieved by locating the turbine closer to a structure is another improvement method of flow control \cite{Coşoiu_2012}. Moreover, constructing an optimized groove contributed to significant improvement in overall performance \cite{seo2016performance} . The improvement was successful because of recovering velocity around the airfoil. It was done on NACA0015 airfoil computationally under 7° angle of attack. Another study was done experimentally on VAWT by introducing leading-edge slot blowing which resulted in 59\% increment in net energy production \cite{sasson2011effect}. Another new concept in a 2016 study discussed dynamic stall control by using co-flow jet on an S809 airfoil numerically; which dramatically increased aerodynamic performance at all angles of attack alleviating extreme aerodynamic load \cite{xu2016dynamic}. Though active flow control may seem quite promising, however, from a purely industrial perspective; it arises higher manufacturing complexity and maintenance; generating higher costs. Hence, passive flow control can be just the right answer. By controlling passive boundary layer control which is achieved by implementing slot on S809 airfoil; the results predict that aerodynamics performance is increased over a specific range of angle of attack, from 10° to 20° angle of attack the airfoil’s lift performance was sufficiently higher than baseline airfoil \cite{BELAMADI201679}. Furthermore, a 2018 study discussed different passive flow control methods including gurney flaps, thin plate at the trailing edge, leading edge micro cylinder, leading-edge serrations, passive movable flaps, tilted blade, J-shaped airfoil, variable droop, airfoil with cavity, adaptive blade, leading edge slat, non-circular gap, flow-deflecting airfoil, vortex generator \cite{zhu2018critical}. The study further concludes that dimple-gurney flap and outboard gurney flap achieve higher lift resulting in higher aerodynamic performance. Lastly, turbine performance improvement is a sufficiently complicated problem that cannot be dealt with in a multi-dimensional problem statement, let alone in a single paper. However, with a critical analysis of a single problem statement, we can clarify several insights that are the main objective of this paper. 

The main objective of this study was to improve the performance of the NACA 6309 airfoil using a plain flap on the trailing edge. Initially, the performance of the NACA 6309 airfoil was studied so that we could compare the effects of the flaps with it. Furthermore, At the trailing edge, upward and downward plain flaps were introduced by generating 15 configurations: 10 configurations for the down plain flap, and 5 for the up plain flap. The flap was added on the downside of the trailing edge from \(0^\circ\) to \(10^\circ\) down position, called the down plain flap whereas the flap was added on the upperside from \(0^\circ\) to \(5^\circ\) upper position, which is called the upper plain flap. The exact location of the trailing edge flap was 70\% position from the leading edge of the airfoil. The lift coefficient, drag coefficient, and lift-to-drag ratio with angle of attack were studied, which helped to determine the best configurations among all. Moreover, an in-depth analysis of velocity and static pressure was analyzed for three airfoils; baseline, best, and worst to find out the framework of the aerodynamic performance.

\section{Simulation Methodology}
\setlength{\parindent}{0.2in}
\hspace{0.2in}
While suspended in the air, an object encounters both lift and drag forces. These forces are encapsulated by the coefficients of lift and drag. These coefficients stand out as the most crucial factors when it comes to comprehending the aerodynamic behavior of an airfoil. The coefficient of lift, illustrated in the first equation, is a dimensionless value in mathematics. It emerges from the interplay of variables such as the object's lift force (\(F_L\)), air density (\(\rho\)), wind speed (\(V\)), and the airfoil's chord length (\(c\)). Similarly, the drag coefficient, calculated using the second equation, is another dimensionless measure representing the effect of drag force (\(F_D\)). The chord length designates the distance between the airfoil's leading and trailing edges. Equally significant, the third equation introduces the pressure coefficient (\(C_p\)), wherein the alteration in pressure \((\Delta P)\) is emphasized. The pressure coefficient serves to convey the variations in pressure across a flow field.

\begin{equation}
C_L = \frac{F_L}{\frac{1}{2} \rho V^2 c}
\end{equation}

\begin{equation}
C_D = \frac{F_D}{\frac{1}{2} \rho V^2 c}
\end{equation}

\begin{equation}
C_p = \frac{\Delta P}{\frac{1}{2} \rho V^2}
\end{equation}

\subsection{CFD Analysis}
\setlength{\parindent}{0.2in}
\hspace{0.2in}
Computational Fluid Dynamics stands out as a prominent methodology for comprehending the fluid dynamics of an airfoil. The foundation of CFD simulations rests upon the Navier-Stokes equation, a descriptor of fluid movement. Rooted in the principles of conservation laws governing fluid properties, this equation's core premise is the preservation of mass, momentum, and energy. In this investigation, we employed ANSYS Fluent as the computational tool. The reliable tool ANSYS Fluent solves the equations conserving mass and momentum to address all flow scenarios. The mass and momentum conservation equations are represented by equations 4 and 5 \cite{tu2023computational}.

\begin{equation}
\frac{\partial \rho}{\partial t} + \nabla \cdot \left( \rho \vec{V} \right) = S_M
\end{equation}

\begin{equation}
\frac{\partial}{\partial t} \left( \rho \vec{V} \right) + \nabla \cdot \left( \rho \vec{V} \vec{V} \right) = -\nabla p + \nabla \cdot \left(\bar{\bar{r}} \right) + \rho \vec{g} + \vec{F}
\end{equation}

\begin{equation}
\bar{\bar{r}} = \mu \left[ \left( \nabla \vec{V} + \nabla \vec{V}^T \right) - \frac{2}{3} \nabla \cdot \vec{V} I\right]
\end{equation}

The conservation of mass equation involves the velocity vector \(\vec{V}\), which depends on both position \((x, y, z)\) and time \((t)\). Additionally, the position coordinates have velocity components \((u, v, w)\), while \(SM\) represents the source term. In the context of the conservation of momentum equation, \(\vec{V}\) again represents the velocity vector discussed earlier. The term \(P\) denotes the static pressure, \(\rho \vec{g}\) and \(\vec{F}\) account for the gravitational body force and external body force, respectively, and finally, \(\bar{\bar{r}}\) denotes the stress tensor. The expression for \(\bar{\bar{r}}\) is illustrated in Equation 6, where viscosity is represented by \(\mu\), and \(I\) stands for the unit tensor.

The numerical formulation of the Navier-Stokes equation used in this study is famously known as the Spalart-Allmaras model. This is a single-equation model that solves a modeled transport equation for kinematic eddy turbulent viscosity \cite{spalart1992one}. It is specifically developed for aerodynamic applications involving wall-bounded flows. Its performance in simulating boundary layers affected by adverse pressure gradients is substantially good. We employed the standard turbulence model, and the formulation of the Spalart-Allmaras model is the transport equation.

\begin{equation}
\frac{D\tilde{u}}{Dt} = P - D + \frac{1}{\sigma} \left[ \nabla \cdot \left( (\mathbf{u} + \tilde{\mathbf{u}}) \nabla \tilde{u} \right) + c_b^2 (\nabla \tilde{u})^2 \right]
\end{equation}

\begin{equation}
P = c_{b1} (1 - f_{t2}) \tilde{S} \tilde{u}
\end{equation}

\begin{equation}
D = \left( c_{w1} f_w - \frac{c_{b1}}{\kappa^2} f_{t2} \right) \left[ \frac{\tilde{u}}{d} \right]^2
\end{equation}

\begin{equation}
f_{t2} = c_{t3} \exp(-c_{t4} \chi^2)
\end{equation}

In Eqn. 7, laminar kinematic viscosity is denoted as \(\nu = \frac{\mu}{\rho}\). Moreover, the production term is given in Eqn. 8 and the wall destruction term is given in Eqn. 9. The equation of laminar suppression term \(f_{t2}\) is given in Eqn. 10, with \(c_{t4}=1.2\) and \(c_{t4}=0.5\). In the original Spalart-Allmaras work, the following values given in Table 1 are considered to be constant in constructing the model. 

\begin{table}[ht]
    \centering
    \caption{Constants for Spalart-Allmaras Model}
    \begin{tabular}{cccccccc}
        \toprule
        Constants & $c_{b1}$ & $c_{b2}$ & $c_{\omega 2}$ & $c_{\omega 3}$ & $\sigma \tilde{v}$ & $c_{v1}$ & $k$ \\
        \midrule
        Value & 0.1355 & 0.622 & 0.31 & 2 & $\frac{2}{3}$ & 7.1 & 0.41 \\
        \bottomrule
    \end{tabular}
\end{table}

\begin{equation}
\frac{\partial \tilde{\nu}}{\partial t} + \frac{\partial}{\partial t} \left( \rho \tilde{\nu} u_i \right) = G_\nu + \frac{1}{\sigma_{\tilde{\nu}}} \left[ \frac{\partial}{\partial x_j} \left\{ \left( \mu + \rho \tilde{\nu} \right) \frac{\partial \tilde{\nu}}{\partial x_j} \right\} + C_{b2} \rho \left( \frac{\partial \tilde{\nu}}{\partial x_j} \right)^2 \right] - Y_\nu + S_{\tilde{\nu}}
\end{equation}

In this equation, \(Y_\nu \) stands for the destruction term and \(G_\nu \) represents the production of turbulent viscosity. Other elements that are used in this procedure are constant.
The Spalart-Allamaras model expects that mesh is sufficiently refined closer to the wall surfaces having non-dimensional wall distance \(y+\) is defined in terms of friction velocity u as denoted in Eqn. 12 and Eqn. 13

\begin{equation}
y^+ = \frac{u_*}{v} 
\end{equation}

\begin{equation}
u_* = \sqrt{\frac{\tau_\omega}{\rho}} 
\end{equation}
The initial pre-processing of the simulation setup is crucial as a one-minute error can result in drastically wrong decisions. Boundary conditions are the most significant initial pre-processors as they direct the flow variables of the designed physical model. Therefore, it is a matter of great attention that assigned boundary conditions namely, inlet, outlet, wall, and interface be consistent. In Table 2, some important boundary conditions are mentioned.

\subsection{Flow Domain and Grid Generation}
\setlength{\parindent}{0.2in}
\hspace{0.2in}
Getting the initial simulation setup right is crucial because even a small mistake could lead to big inaccuracies in conclusions. Boundary conditions are especially important in the beginning, as they control how the flow behaves in the model. So, making sure that the assigned conditions at the inlet, outflow, wall, and interface matchup is really important. Initial boundary conditions are presented in Table 1.

\begin{table}[h]
    \centering
    \caption{Initial Boundary Conditions of the CFD Analysis}
    \label{tab:multiples3}
    \begin{tabular}{>{\raggedright\arraybackslash}p{0.08\linewidth}>{\raggedright\arraybackslash}p{0.1\linewidth}>{\raggedright\arraybackslash}p{0.08\linewidth}> {\raggedright\arraybackslash}p{0.1\linewidth}>{\raggedright\arraybackslash}p{0.12\linewidth}>{\raggedright\arraybackslash}p{0.12\linewidth}>{\raggedright\arraybackslash}p{0.1\linewidth}>{\raggedright\arraybackslash}p{0.08\linewidth}}
    \toprule 
    Simulation Property & Parameters & Solver Type & Time & Viscous Model & Number of Iterations & Momentum & Pressure Velocity Coupling \\  
    & Value & Pressure based  & Steady & Spalart-Allmaras & Close to 300 & Second-order upwind & Simple \\
    \midrule
    Fluid Property & Parameters & Fluid & Density (kg/m\(^3\)) & Viscosity (kg/m\(\cdot\)s) & Angle of Attack & Reynold Number & Pressure \\  
    & Value & Air & 1.225 & 1.7894 & \(0^\circ\) to \(20^\circ\) & 300000 & 1 atm \\
    \bottomrule
    \end{tabular}
\end{table}

In our study, C-type mesh, which is preferable for airfoil simulation, was utilized. The semicircle radius was 10 and the width of the rectangle was 10. The wall condition for the airfoil was not taken into account. Reynolds number of 1 million was previously mentioned, and using this value, we calculated the inlet velocity to be 14.82 ms-1. In order to find a smooth mesh, refinement was utilized. The value of \(y+\) was kept under control in order to maintain the accuracy of the results. In our mesh, the node number was 85,000 whereas the element number was 0.17 million.
\begin{figure}[h]
    \centering
    \begin{subfigure}[b]{0.31\textwidth}
        \centering
      \includegraphics[width=\textwidth]{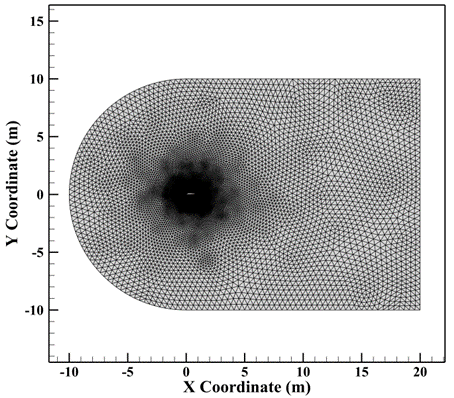}
        \caption{}
        \label{}
    \end{subfigure}
    \hfill
    \begin{subfigure}[b]{0.31\textwidth}
        \centering
        \includegraphics[width=\textwidth]{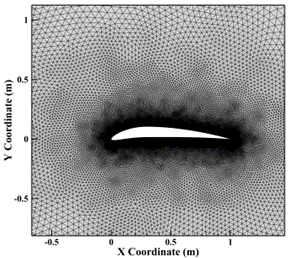}
        \caption{}
        \label{}
  \end{subfigure}
  \hfill
    \begin{subfigure}[b]{0.31\textwidth}
        \centering
        \includegraphics[width=\textwidth]{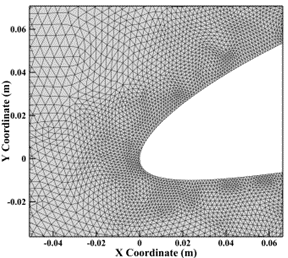}
        \caption{}
        \label{}
  \end{subfigure}
    \caption{\small Mesh of C-type computational domain constructed using ANSYS. (a) Airfoil (b)Leading edge (c) Trailing edge.}
    \label{fig:sidebyside}
    
\end{figure}

\subsection{Validation of CFD Model}
\setlength{\parindent}{0.2in}
\hspace{0.2in}
In order to measure the accuracy of the turbulence model and the overall generated result of the paper, we have compared lift coefficient curves with experimental and numerical data developed by Eleni (Douvi C. Eleni, 2012). In the research work, the Reynolds number was 30,00,000 at 300K temperature having 1.225 kg/m3 density and 1.8×10-5 Ns-m-2. On NACA 0012 airfoil from -\(12^\circ\) to 20° angle of attack was experimented and simulated. Abbott's (Abbott, 1945)  experiments and our results provide further support for the study. In this work, we used numerical techniques to carefully assess the experimental data. Our results' degree of agreement is shown in Figure 2. The experimental and numerical results agreed well, with negligible error. This enhances our study's dependability and offers a strong foundation for more investigation.

\subsection{Airfoil Conceptualization}
\setlength{\parindent}{0.2in}
\hspace{0.2in}
In our work, NACA 6309 airfoil was selected, which is preferable for horizontal-axis wind turbines blade. NACA airfoils are designed by National Advisory Committee for Aeronautics(NACA) and denoted by NACAYYXX, where the first digit denotes the percentage of maximum camber with respect to its chord, the second digit represents the position of that maximum camber from the leading edge in tenths of the chord. In addition, the maximum thickness-to-chord ratio is denoted by the last two digits. A 1000mm chord model of NACA 6309 airfoil was used for this analysis with 100 different coordinate points. The airfoil was generated using the following polynomial:

\begin{equation}
y_t = 5t \left[ 0.2969\sqrt{\frac{x}{c}} - 0.1260\left(\frac{x}{c}\right) - 0.31516\left(\frac{x}{c}\right)^2 + 0.2843\left(\frac{x}{c}\right)^3 - 0.1015\left(\frac{x}{c}\right)^4 \right]
\end{equation}

\begin{figure}[t]
    \centering
      \includegraphics[width=0.49\textwidth]{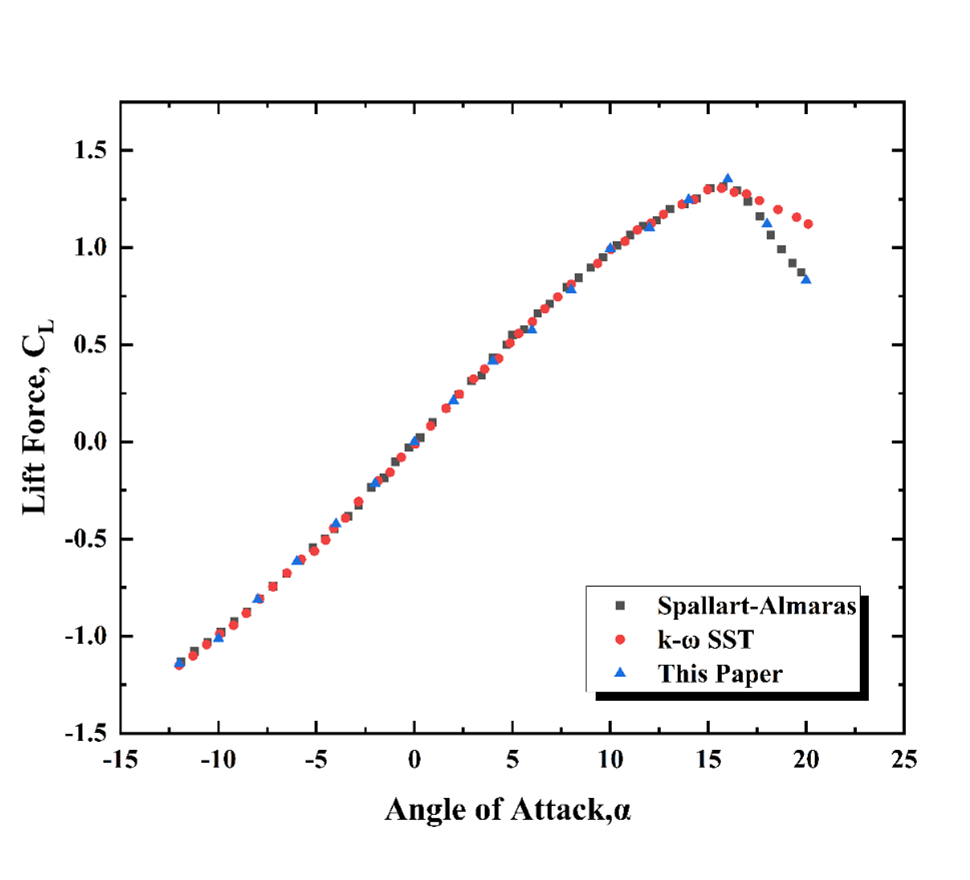}
        \caption{Validation of the simulated results according to Lift.}
        \label{}
\end{figure}

Here, c indicates the chord length, and x is the horizontal coordinate, which stays between the starting from the leading edge (0) to the trailing edge \((c)\). Furthermore, \(y_t\) is the half thickness at a given value of \(x\), and t is the maximum thickness as a fraction of the chord. On the trailing edge, a minute gap was induced to enhance the performance. In this regard, a plain flap was used in the trailing edge. The exact position of the flap was at 0.7c (700mm) position from the leading edge. In other words, the plain flap position was 300 mm from the trailing edge. A flap is an important section that is mainly incorporated in airfoil to enhance the lift effect along with the reduction of stalling speed. In addition, it helps to minimize the fluctuation of load at higher wind speed which results in an improved life-cycle. A total of sixteen configurations—five upward plain flaps from \(1^\circ\) to \(5^\circ\), ten downward plain flaps from \(1^\circ\) to \(10^\circ\), and a baseline airfoil (where no flap is utilized)—were examined in this study.

\section{Results And Discussion}
\setlength{\parindent}{0.2in}
\hspace{0.2in}
The results of NACA6309 under 16 different conditions (baseline, \(1^\circ\) to \(10^\circ\) plain down flap, and \(1^\circ\) to \(5^\circ\) plain up flap) are presented in this section to find out the best-performing airfoil, along with the worst. The best and worst performing airfoils are compared with baseline airfoils in order to comprehend the actual improvement of such performance. To begin with, the airfoils’ lift coefficient with the angle of attack is discussed. Furthermore, the coefficient of drag is introduced. Following that, the ratio of lift and drag is analyzed to see which airfoils performed the best and worst. Then, to find out the underlying mechanism of such performance, static pressure contours are developed. And finally, the velocity contours are presented for the three airfoils (best, baseline, and worst).
 
\subsection{Impact of Lift Coefficient (\(C_L\)) With Angle of Attack (AOA)}
\setlength{\parindent}{0.2in}
\hspace{0.2in}
The airfoil’s lift and drag forces are primarily influenced by the angle of attack. When the angle of attack increases from 0°, lift performance increases up to a certain angle. After a certain angle, the lift force decreases whereas the drag force increases. This certain point is called the stall angle. The stall angle is the main reason for the transition from laminar to turbulence around the airfoils. With a different angle of attack with respect to airfoil performance ratio which lift to drag ratio is presented in Fig. 3. In the baseline NACA6309 airfoil, the lift performance increases with the angle of attack until \(13^\circ\), after that stall occurs; from 0° to \(13^\circ\) the lift increases linearly. The lift at 0° AOA is 0.5914 and the maximum generated lift is 1.6054. The down-flap from \(1^\circ\) to \(10^\circ\) gave a higher initial lift in all cases, whereas in the up-flap portion, all the initial lift performances were poorer than the baseline airfoil. 

\begin{figure}[h]
    \centering
      \includegraphics[width=0.49\textwidth]{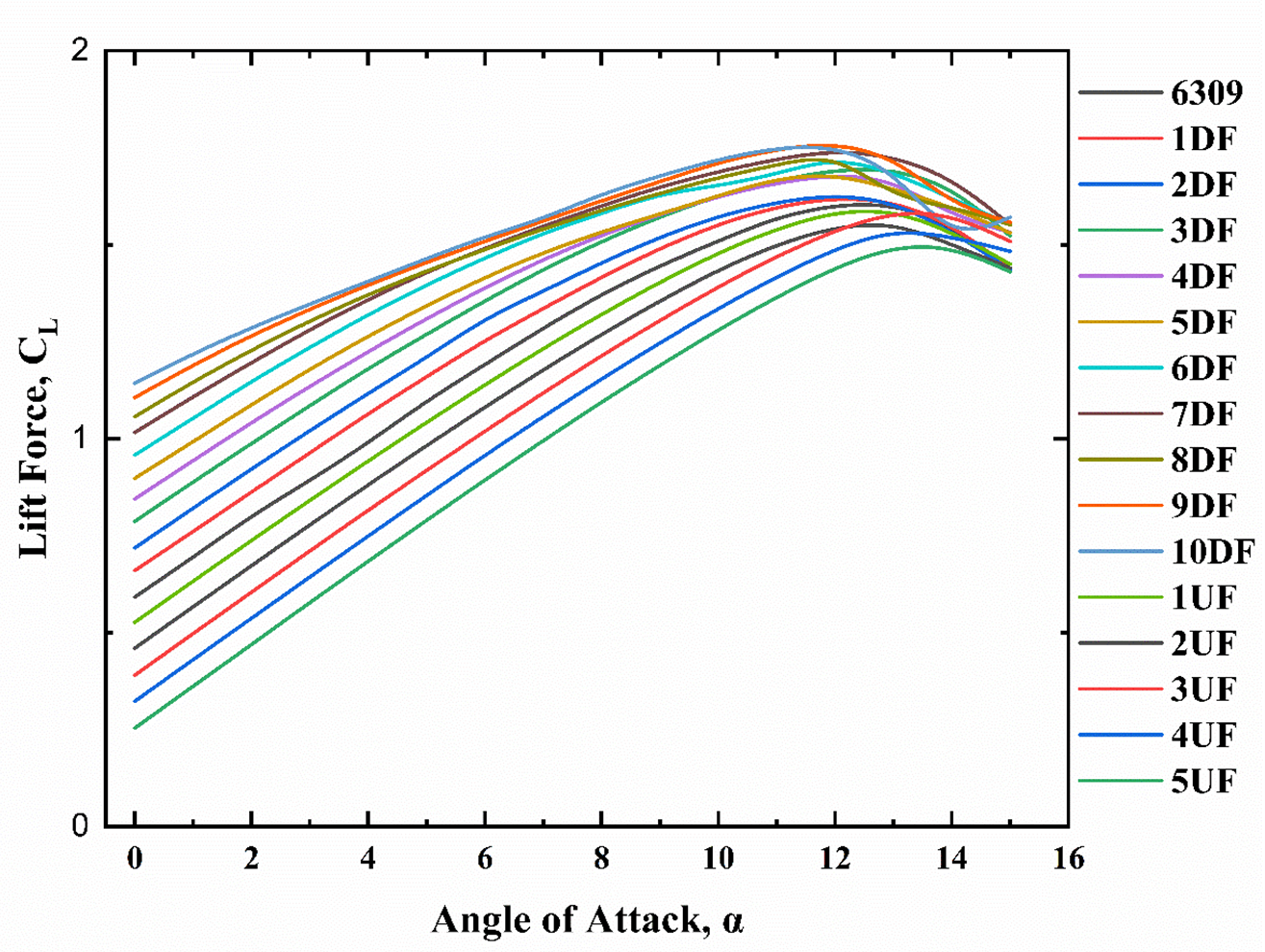}
        \caption{Impact of lift on the studied airfoils with angles of attack}
        \label{}
\end{figure}
The lift performance dramatically increases with increasing down-flap angle, even the initial lift performance happened to be increasing with increasing angle of down-flap, which is apparent from Fig. 4. At \(1^\circ\) and 2°down-flap, the stall occurs at \(12^\circ\) angle of attack, until the lift increases linearly. From \(4^\circ\) down-flap to \(9^\circ\) down-flap, the stall angle remains \(12^\circ\) and the performance was better with increasing down-flap angle. At the \(10^\circ\) down-flap the lift performance outperformed all the airfoils, with stall occurring at \(12^\circ\), nevertheless after stalling the lift performance remained sufficiently greater than average. On the opposite side, the up-flap lift force initially performed poorer than average but with increasing angle of attack, it catches up with most of the higher-performing airfoils.

In \(1^\circ\)up-flap, the stall occurs at \(13^\circ\) angle of attack, still the recovering lift results were good. At \(3^\circ\)up-flap the lift force acts similarly compared to \(1^\circ\) up-flap. As, both the up-flap and down-flap airfoils performed sufficiently well, the main deciding factor of performance will be the ratio of lift and drag. 
\begin{figure}[h]
    \centering
      \includegraphics[width=0.49\textwidth]{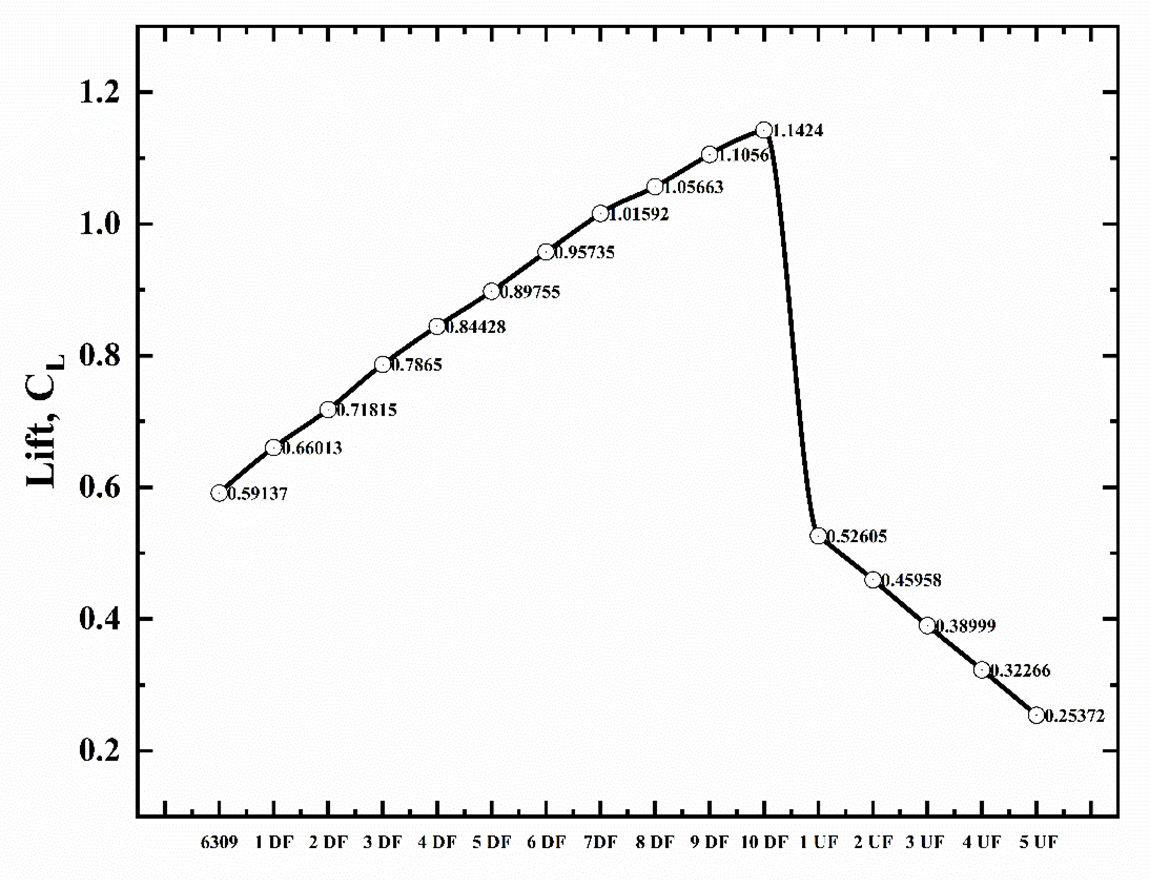}
        \caption{Comparison of lift with down and up flap at zero-degree angle of attack.}
        \label{}
\end{figure}
When the plain down flap is utilized in the trailing edge of the NACA 6309 airfoil, an augmentation of the lift coefficient is noticed. For instance, when \(2^\circ\) plain down flap is utilized in the trailing edge of the airfoil, the lift coefficient is 0.72 at \(0^\circ\) angle of attack. Following the same trend discussed before, the value of \(C_L\) enhances till the stall angle which is \(12^\circ\). The lift value at the stall angle is 1.63, which is higher compared to the baseline airfoil. When the down flap degree is enhanced, especially for \(6^\circ\) to \(10^\circ\) plain down flap configurations, the value of \(C_L\) is high at almost every AOA compared to the up flap and baseline configurations. Especially, at \(10^\circ\) plain down flap configuration, the coefficient of lift is high at every angle of attack compared to the other conditions. However, while using the up flap at the trailing edge of the NACA 6309 airfoil, the increment of the lift coefficient is not adequate in every case. Although, \(1^\circ\) and \(2^\circ\) up flap denotes a relatively moderate result, \(2^\circ\),\(3^\circ\) and \(5^\circ\) degree up flap shows a weaker \(C_L\) vs. AOA curve compared to the baseline airfoil.
\begin{figure}[h]
    \centering
      \includegraphics[width=0.49\textwidth]{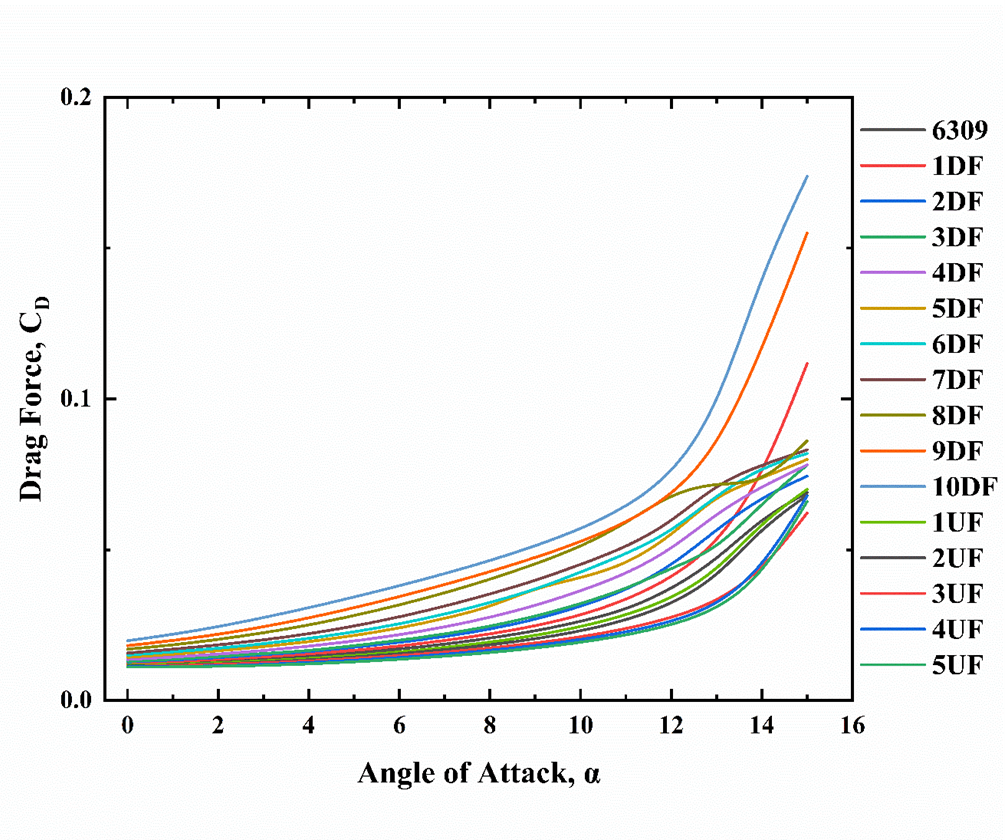}
        \caption{Impact of Drag on the studied airfoil with respect to angle of attack.}
        \label{}
\end{figure}
\subsection{Impact of Drag Coefficient (\(C_D\)) with Angle of Attack (AOA)}
\setlength{\parindent}{0.2in}
\hspace{0.2in}
The value of the drag coefficient (\(C_D\)) is obligatory, as it helps to understand the aerodynamic performance of the airfoil appropriately. The drag coefficient arises from mainly two actions: the pressure difference between the leading and trailing edge of the airfoil and the viscous resistance coming from the surface of the airfoil; the first one is pressure drag and the second one is viscous or friction drag. Between the two, the major portion of the drag coefficient is caused by pressure which deteriorates the total performance. 
\begin{figure}[h]
    \centering
      \includegraphics[width=0.49\textwidth]{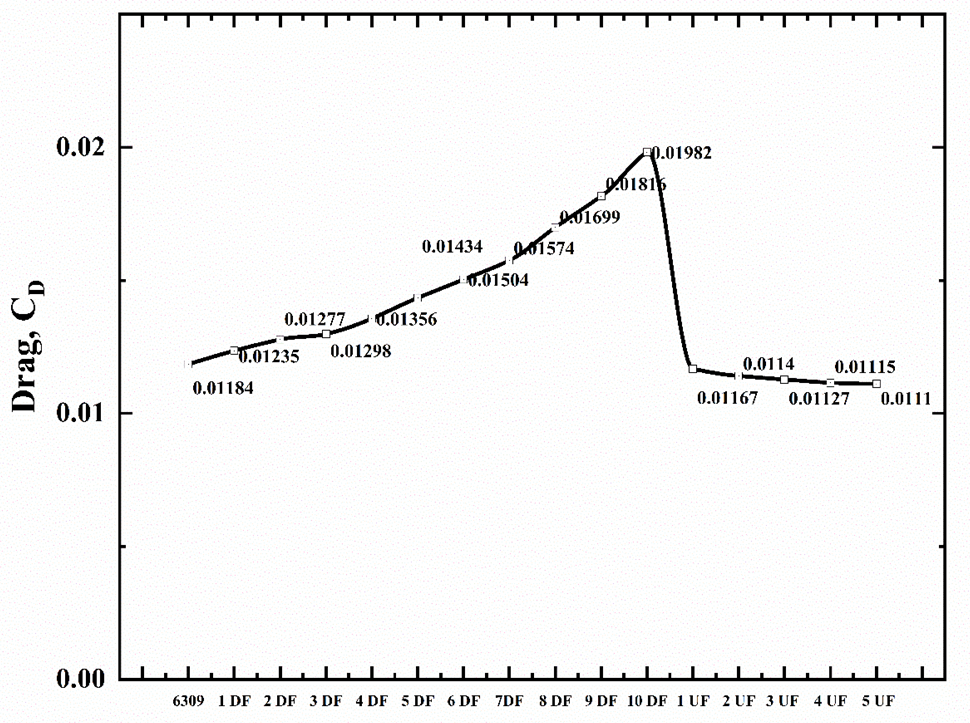}
        \caption{Comparison of Drag with down and up flap at the zero-degree angle of attack.}
        \label{}
\end{figure}

In spite of the fact that the value of the drag coefficient remains within the acceptable limit at the lower angle of attack (AOA), the value starts to enhance dramatically at the higher AOA. The scenario is the almost same in every case of our study. In the baseline NACA6309 airfoil, the drag coefficient is 0.0118 at \(0^\circ\) AOA. This value enhances steadily until it touches \(10^\circ\), where the value of \(C_D\) is 0.0262. After crossing this stage, a dramatic enhancement is noticed. At \(15^\circ\) AOA, this value is 0.0689, which indicates a huge increment of the drag coefficient. The more the down-flap angle increases, especially from \(6^\circ\) to \(10^\circ\) plain down-flap configurations, the poorer the airfoil performed drag-wise. For instance, while using \(10^\circ\) down-flap at the trailing edge, the value of the \(C_D\) is the maximum in every AOA compared to the other conditions. In detail, In \(10^\circ\) down-flap the drag coefficient is maximum at \(15^\circ\) AOA has a more than 700\% increment from \(0^\circ\) angle of attack. But from Fig. 6, it is evident that the drag coefficient is incredibly lower at up-flap conditions. In the \(1^\circ\)up-flap configuration, the drag coefficient at \(0^\circ\) angle of attack is 1.44\% lower than the baseline airfoil and almost 70\% lower than \(10^\circ\) down-flap configuration. At the maximum angle of attack, the drag coefficient is 0.07, with an increment of 4.9\% from \(0^\circ\) angle of attack. Throughout the up-flap configuration, the drag coefficient remains relatively stable compared to the down-flap conditions.

\subsection{Impact of Lift-Drag Ratio with Angle of Attack}
\setlength{\parindent}{0.2in}
\hspace{0.2in}
In the previous sections, the coefficient of lift and drag are discussed with the angle of attack. However, the coefficient of lift to drag ratio provides a clear indication of which condition the airfoil will be the most efficient. The plain down-flap configurations provide a higher lift coefficient with increasing angle of the attack, whereas, the plain up-flap configurations performed relatively unsatisfactory. In addition, with increasing down-flap angle the drag performance worsened. On the other hand, the up-flap configuration provides the best drag performance result, meaning the drag was sufficiently lower than the baseline and down-flap. As a conflict of performance arises, therefore only the ratio of lift to drag can give the proper insight into the overall performance of airfoils.
\begin{figure}[h]
    \centering
      \includegraphics[width=0.55\textwidth]{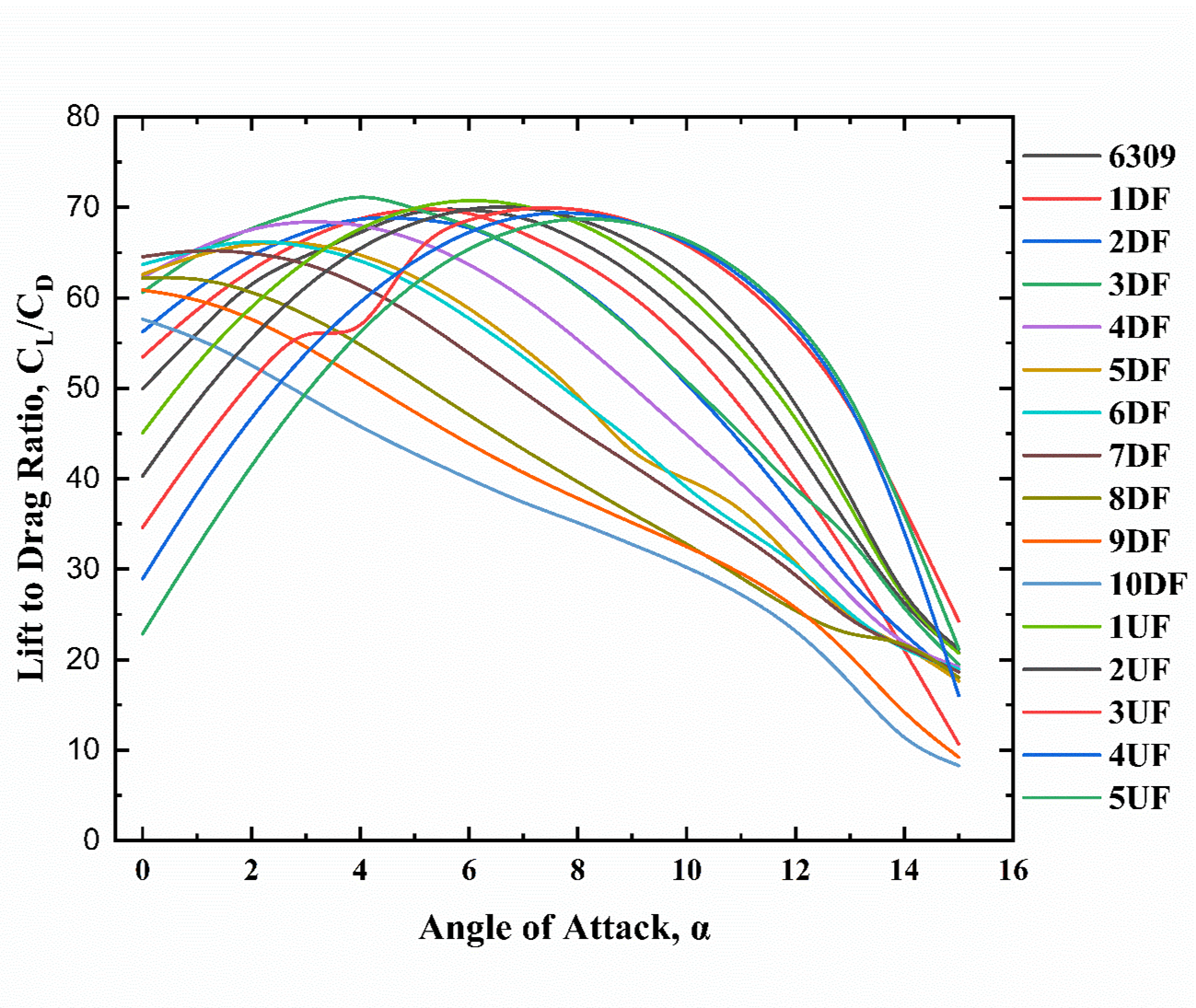}
        \caption{Comparison of Lift to Drag Ratio with Angle of Attack}
        \label{}
\end{figure}
The optimum airfoil design would be the airfoil that has the highest lift with the lowest drag which will effectively give the highest achievable power generation for such wind turbine. This article will find the best one which is a relatively higher lift performer with a relatively lower drag performing airfoil resulting in maximization of power generation. Intriguingly, in every case, the airfoil is following the same trend. Initially, the value of \(C_L\) and \(C_D\) ratio enhances to \(6^\circ\) to \(7^\circ\), however, this value starts to fall after crossing this angle of attack. At \(10^\circ\) down-flap, the lift-to-drag ratio is the worst-performing airfoil. Although its lift performance is the best, it can be seen from the previous section that the drag coefficient increases drastically leaving the airfoil performance in derail. The lift-to-drag ratio at \(2^\circ\) plain down flap configuration is adequate at every AOA, however, it is not the optimal one. On the contrary, in the up-flap configurations, the performance is the best. In \(1^\circ\) up-flap, the lift-to-drag ratio with respect to the angle of attack is the highest among all, and the maximum value was 71.112at \(7^\circ\) AOA. Though the lift performance is relatively poorer in this airfoil, with drag recovery; the overall performance outperforms the down-flap configurations. Moreover, the other up-flap configurations are comparatively higher performers than their down-flap counterparts. But with increasing up-flap angle the performance decreases too. Therefore, it can be stated, with increasing down-flap and up-flap angles the performance decreases and \(1^\circ\) up-flap is the highest power-producing airfoil configuration.  

\subsection{Investigation of Static Pressure}
\setlength{\parindent}{0.2in}
\hspace{0.2in}
The wings of any aerodynamic body create lift by generating higher pressure developed at the lower part of the body than the upper part, creating a net upward force resulting in power generation in wind turbines. Now if the pressure below the airfoil gets increased with a higher angle of attack, the net force increases creating higher lift but also overall drag increases too. Drag development increases when the pressure at the leading edge of the airfoil is higher than at the trailing edge. From the zero-degree angle of attack diminutive drag force development is obvious, but we need to find the optimum airfoil design (in our case the one-degree up-flap) where the overall drag-lift performance gets better. 
\begin{figure}[htbp]
    \centering
    \begin{subfigure}[h]{0.31\textwidth}
        \centering
      \includegraphics[width=\textwidth]{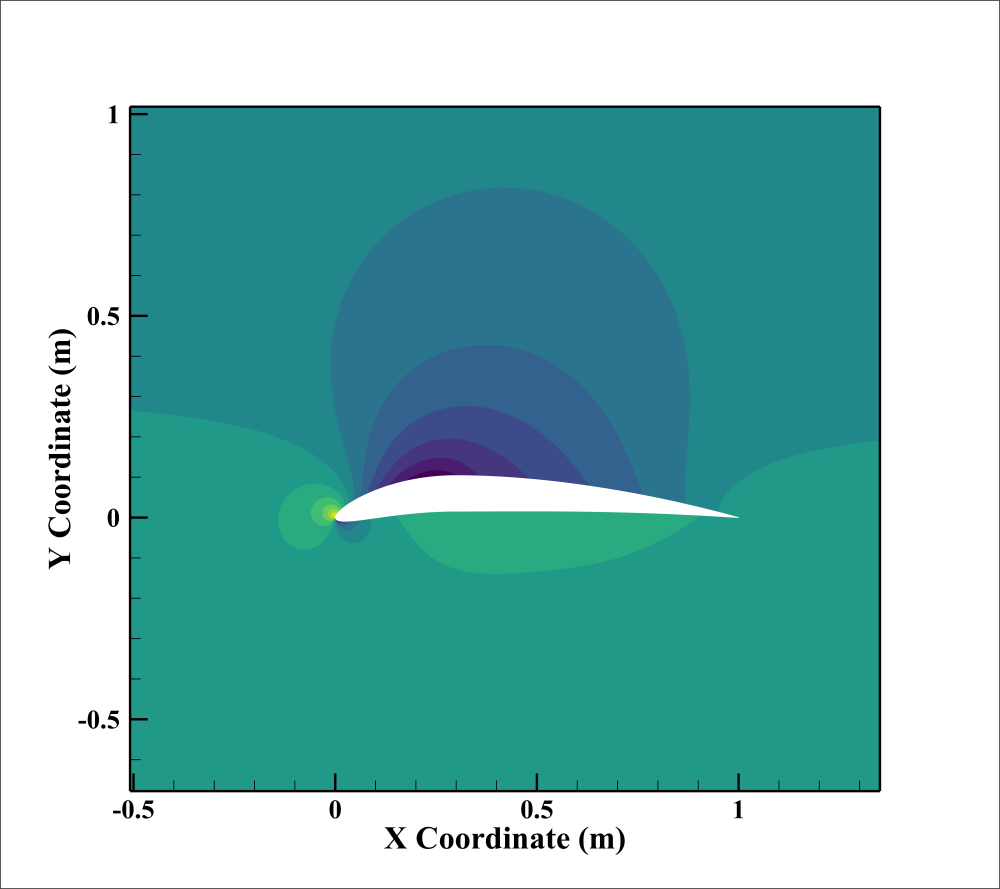}
        \caption{\(0^\circ\)}
        \label{}
    \end{subfigure}
    \hfill
    \begin{subfigure}[h]{0.31\textwidth}
        \centering
        \includegraphics[width=\textwidth]{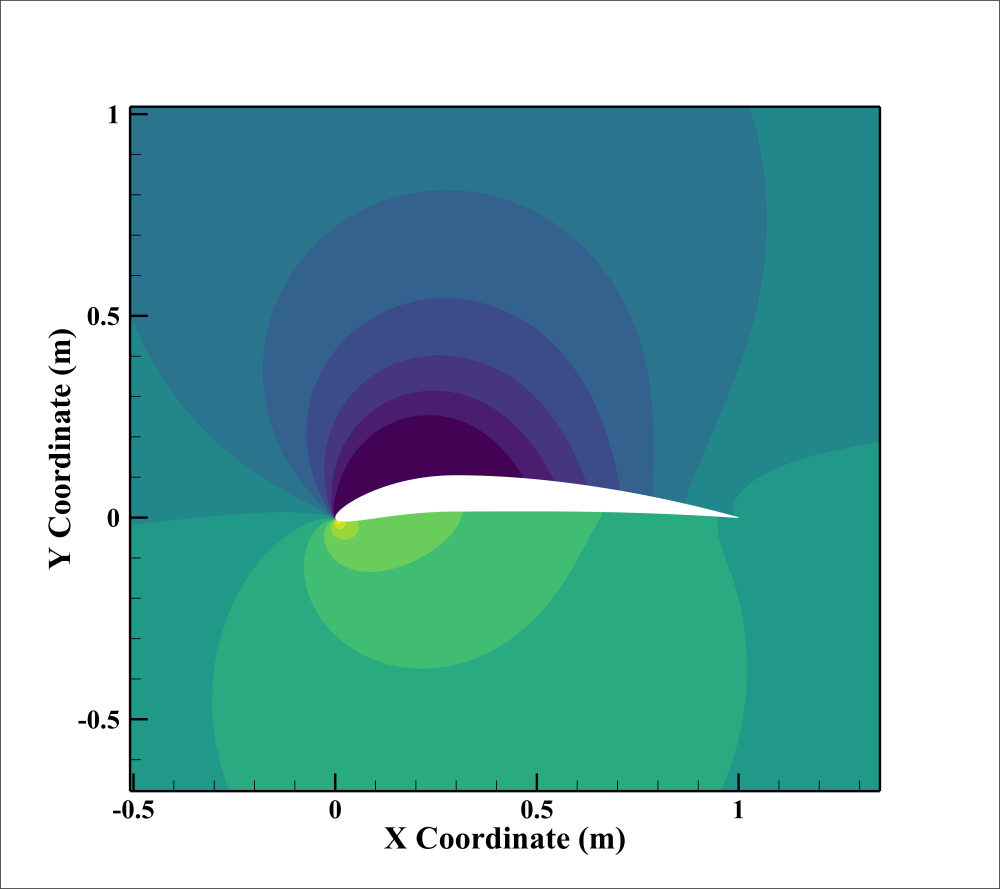}
        \caption{\(7^\circ\)}
        \label{}
  \end{subfigure}
  \hfill
    \begin{subfigure}[h]{0.31\textwidth}
        \centering
        \includegraphics[width=\textwidth]{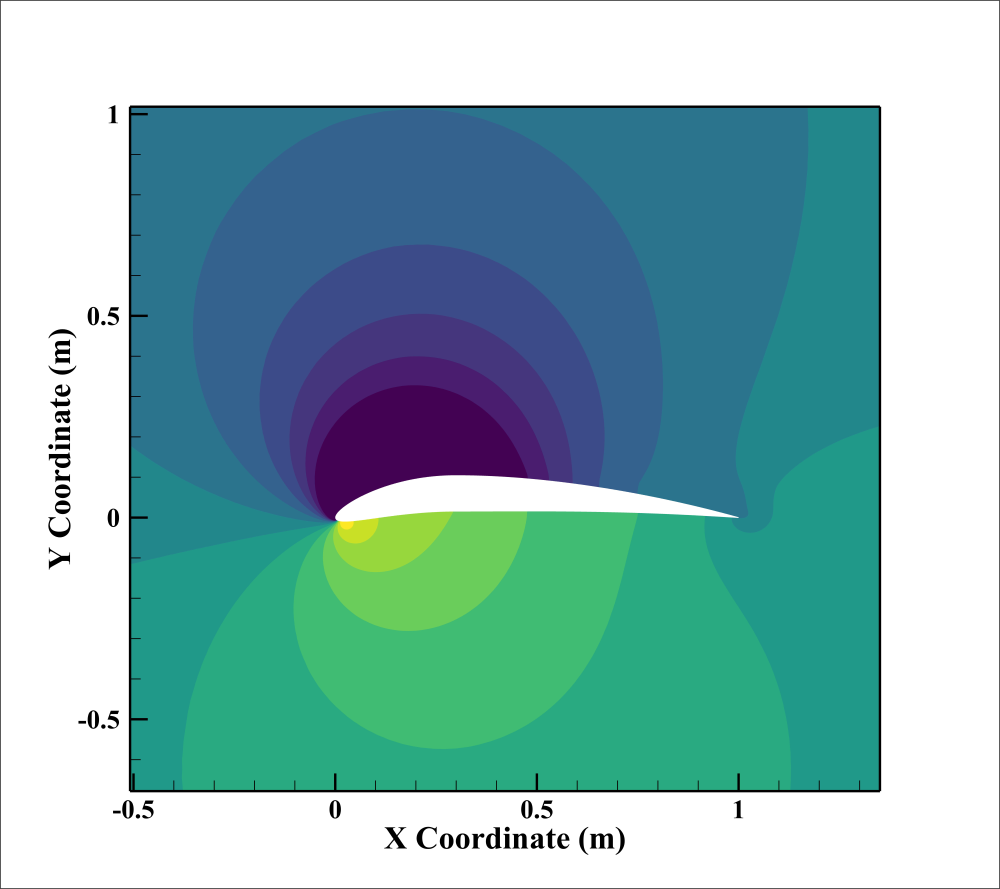}
        \caption{\(12^\circ\)}
        \label{}
  \end{subfigure}
  \hfill
  \vspace{5em}
    \begin{subfigure}[h]{0.31\textwidth}
        \centering
        \includegraphics[width=\textwidth]{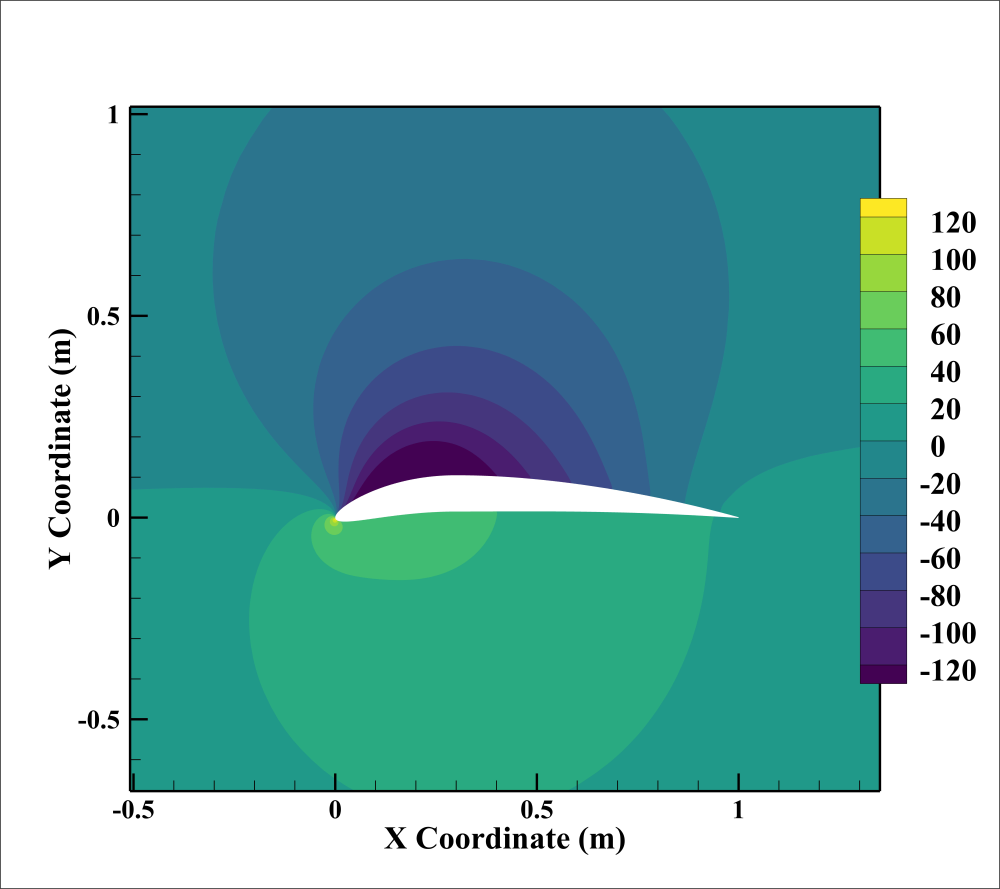}
        \caption{\(4^\circ\)}
        \label{}
  \end{subfigure}
  \hfill
    \begin{subfigure}[h]{0.31\textwidth}
        \centering
        \includegraphics[width=\textwidth]{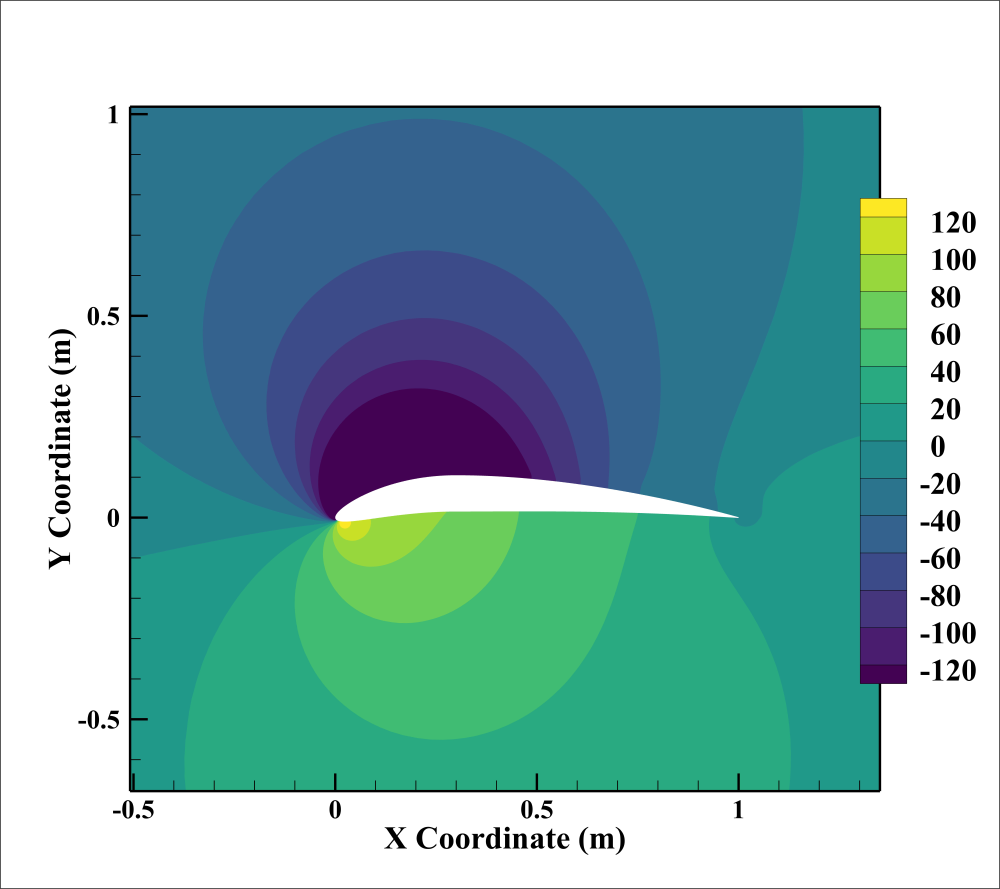}
        \caption{\(11^\circ\)}
        \label{}
  \end{subfigure}
  \hfill
    \begin{subfigure}[h]{0.31\textwidth}
        \centering
        \includegraphics[width=\textwidth]{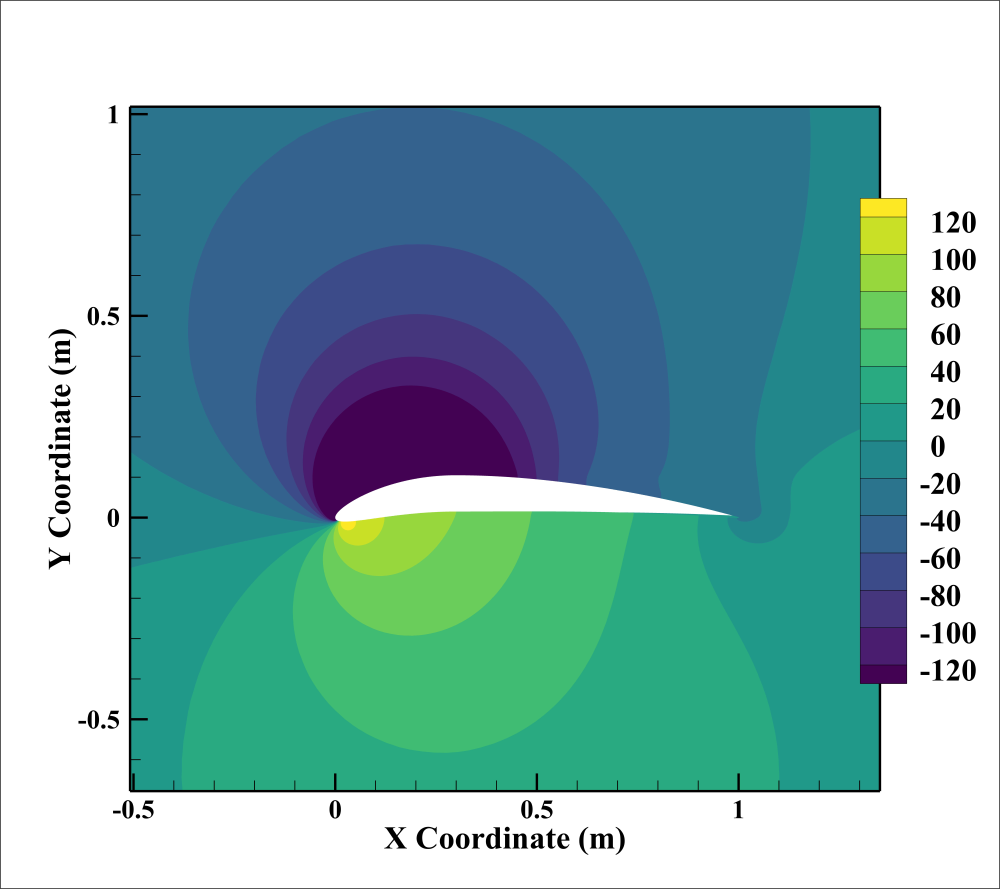}
        \caption{\(13^\circ\)}
        \label{}
  \end{subfigure}
    \caption{\small NACA 6309 pressure contours at \(0^\circ\), \(4^\circ\), \(7^\circ\), \(12^\circ\), \(13^\circ\) angle of attack.}
    \label{fig:sidebyside}
\end{figure}
\begin{figure}[htbp]
    \centering
    \begin{subfigure}[b]{\textwidth}
        \centering
        \begin{subfigure}[b]{0.31\textwidth}
            \includegraphics[width=\textwidth]{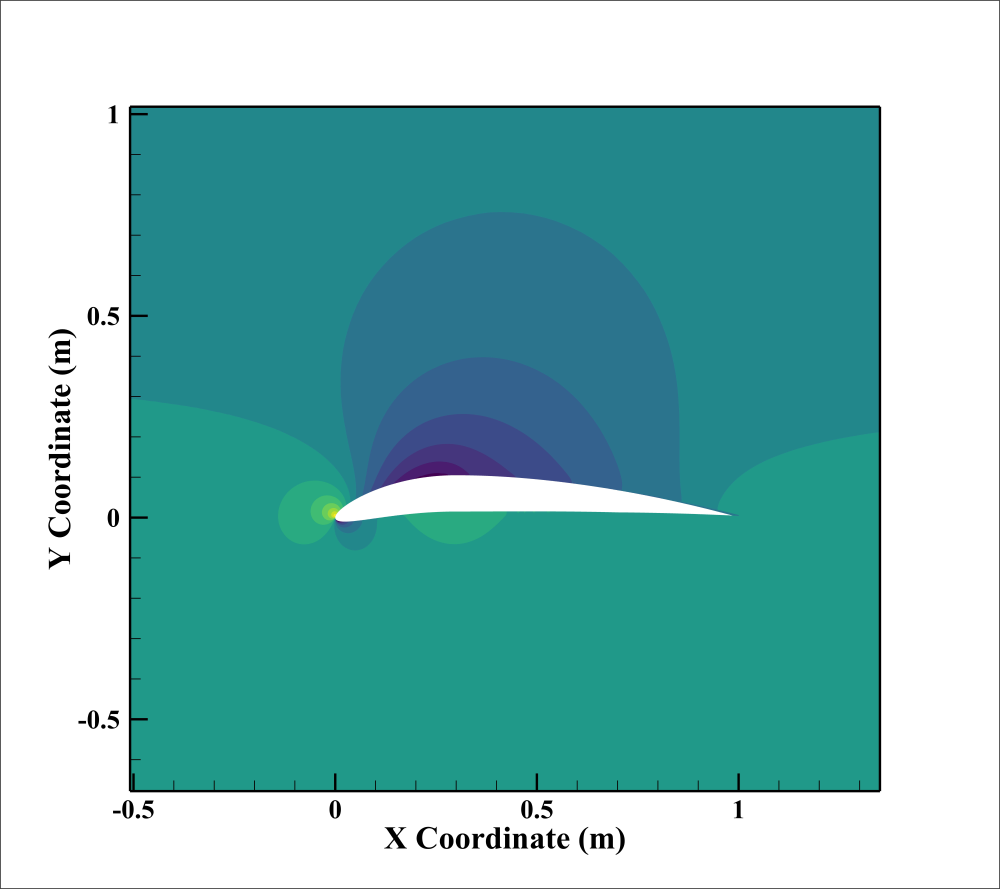}
            \caption{\(0^\circ\)}
        \end{subfigure}
        \hfill
        \begin{subfigure}[b]{0.31\textwidth}
            \includegraphics[width=\textwidth]{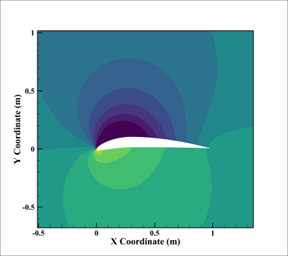}
            \caption{\(7^\circ\)}
        \end{subfigure}
        \hfill
        \begin{subfigure}[b]{0.31\textwidth}
            \includegraphics[width=\textwidth]{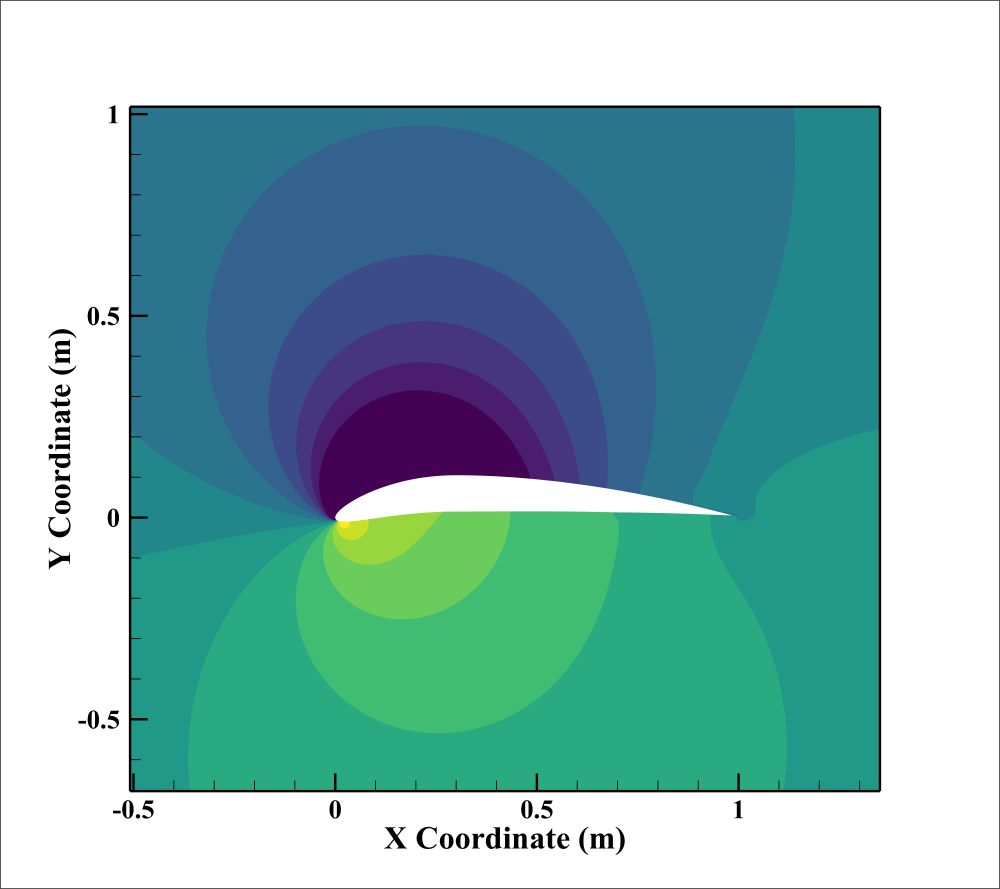}
            \caption{\(12^\circ\)}
        \end{subfigure}
        
        \vspace{1em}
        
        \begin{subfigure}[b]{0.31\textwidth}
            \includegraphics[width=\textwidth]{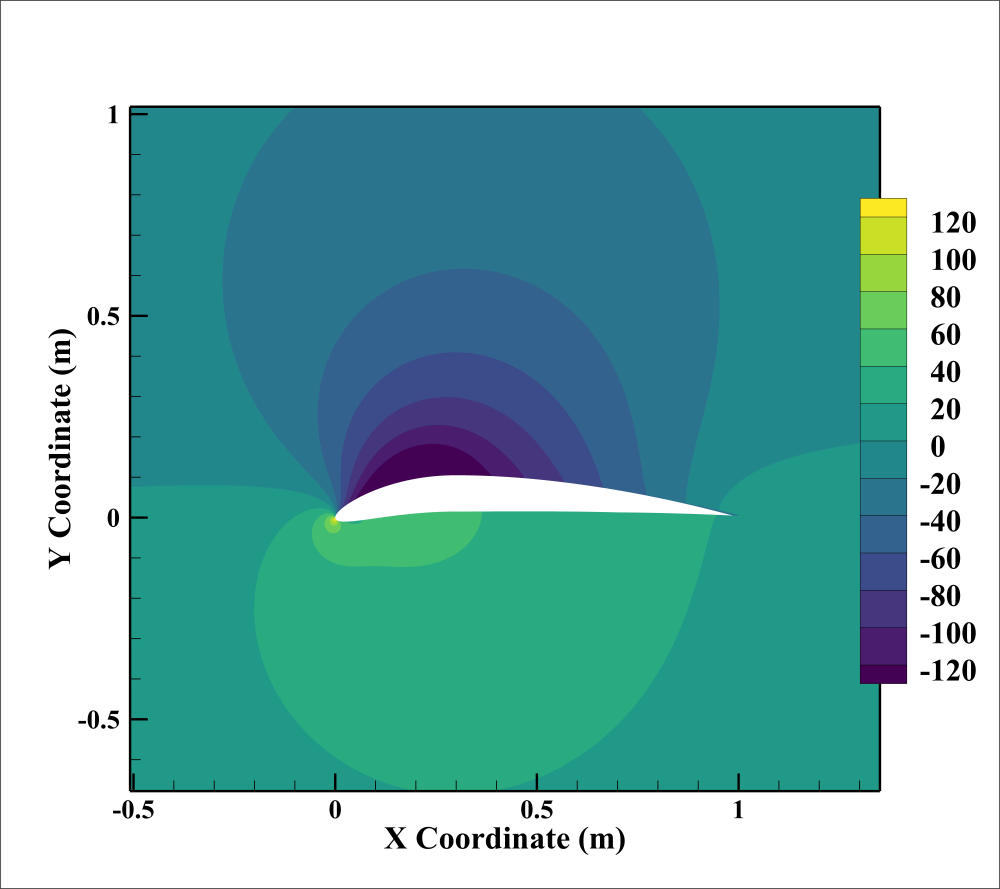}
            \caption{\(4^\circ\)}
        \end{subfigure}
        \hfill
        \begin{subfigure}[b]{0.31\textwidth}
            \includegraphics[width=\textwidth]{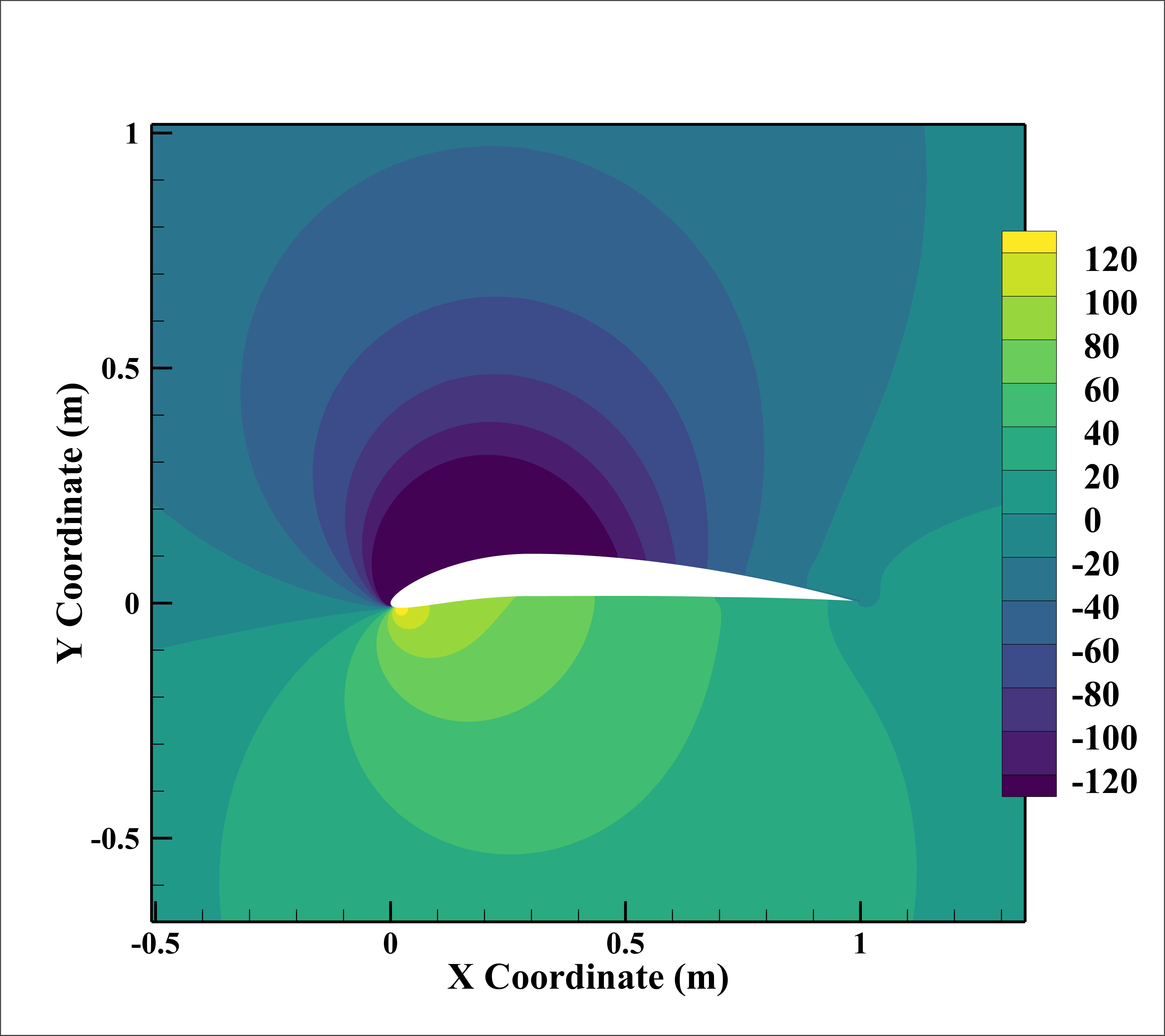}
            \caption{\(11^\circ\)}
        \end{subfigure}
        \hfill
        \begin{subfigure}[b]{0.31\textwidth}
            \includegraphics[width=\textwidth]{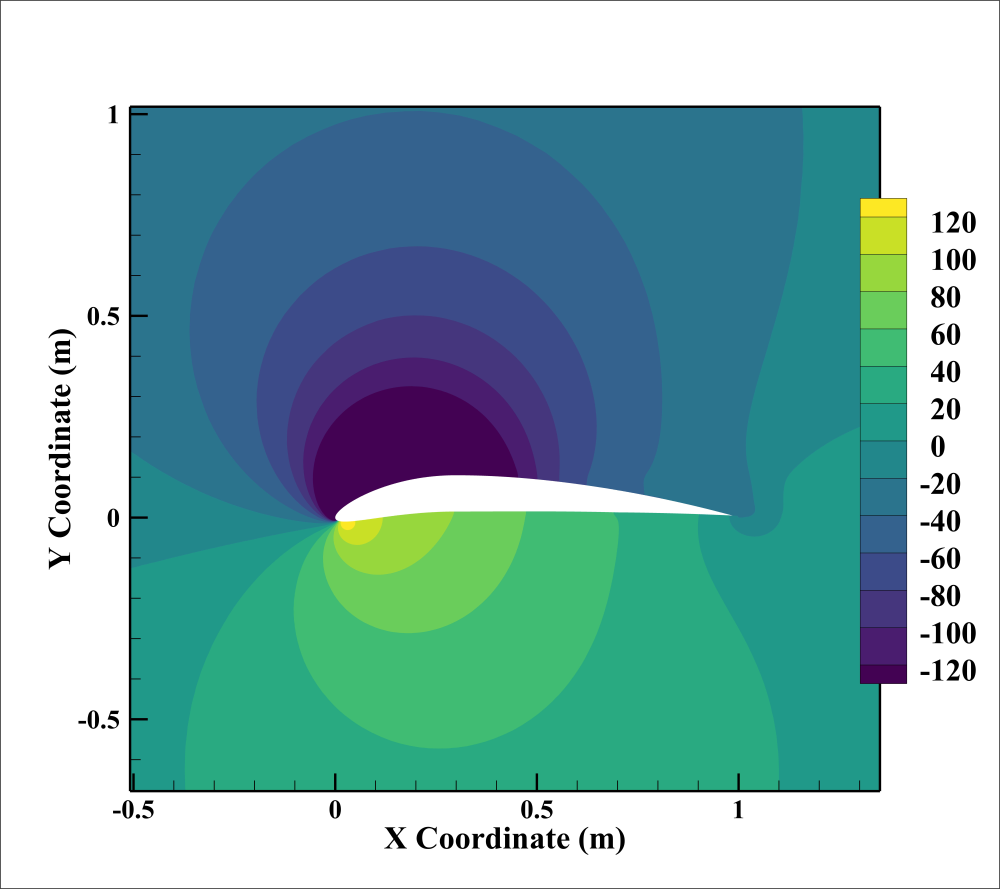}
            \caption{\(13^\circ\)}
        \end{subfigure}
    \end{subfigure}
   \caption{\small Modified NACA6309 with 1 Degree Up-flap Pressure Contour at  \(0^\circ\), \(4^\circ\), \(7^\circ\), \(12^\circ\), \(13^\circ\) angle of attack.}
    \vspace{2em} 

    \setcounter{subfigure}{0}

    \begin{subfigure}[b]{\textwidth}
        \centering
        \begin{subfigure}[b]{0.31\textwidth}
            \includegraphics[width=\textwidth]{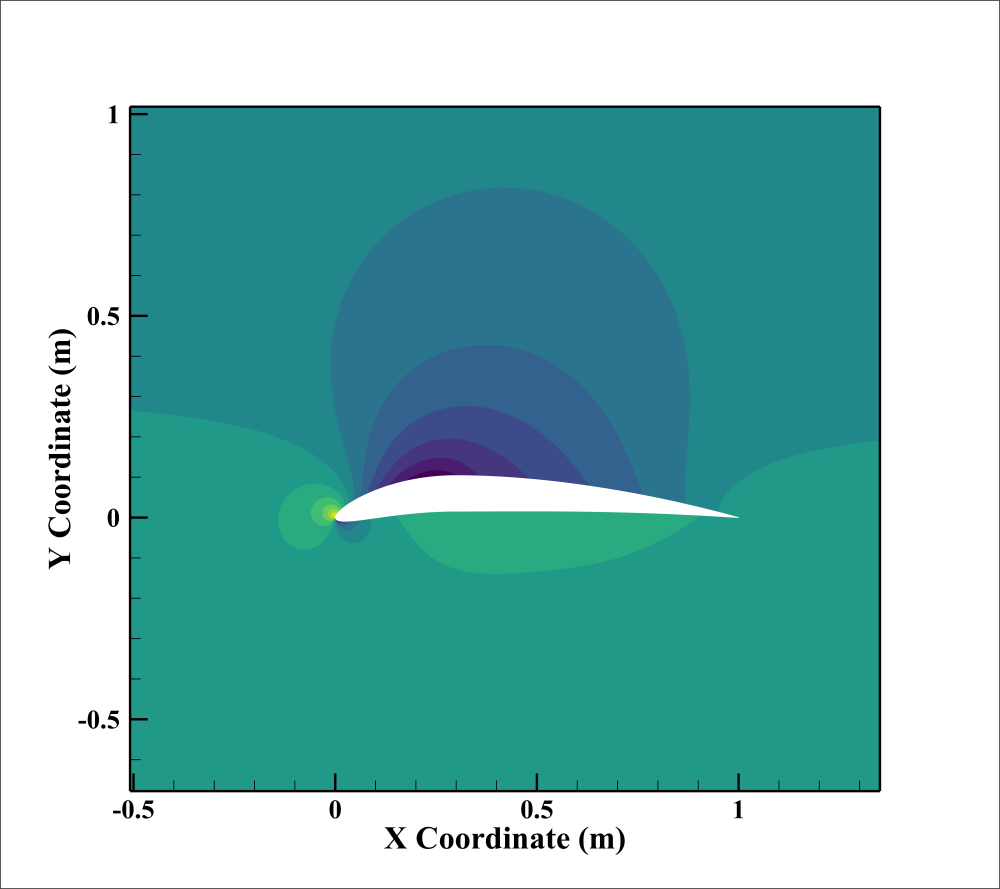}
            \caption{\(0^\circ\)}
        \end{subfigure}
        \hfill
        \begin{subfigure}[b]{0.31\textwidth}
            \includegraphics[width=\textwidth]{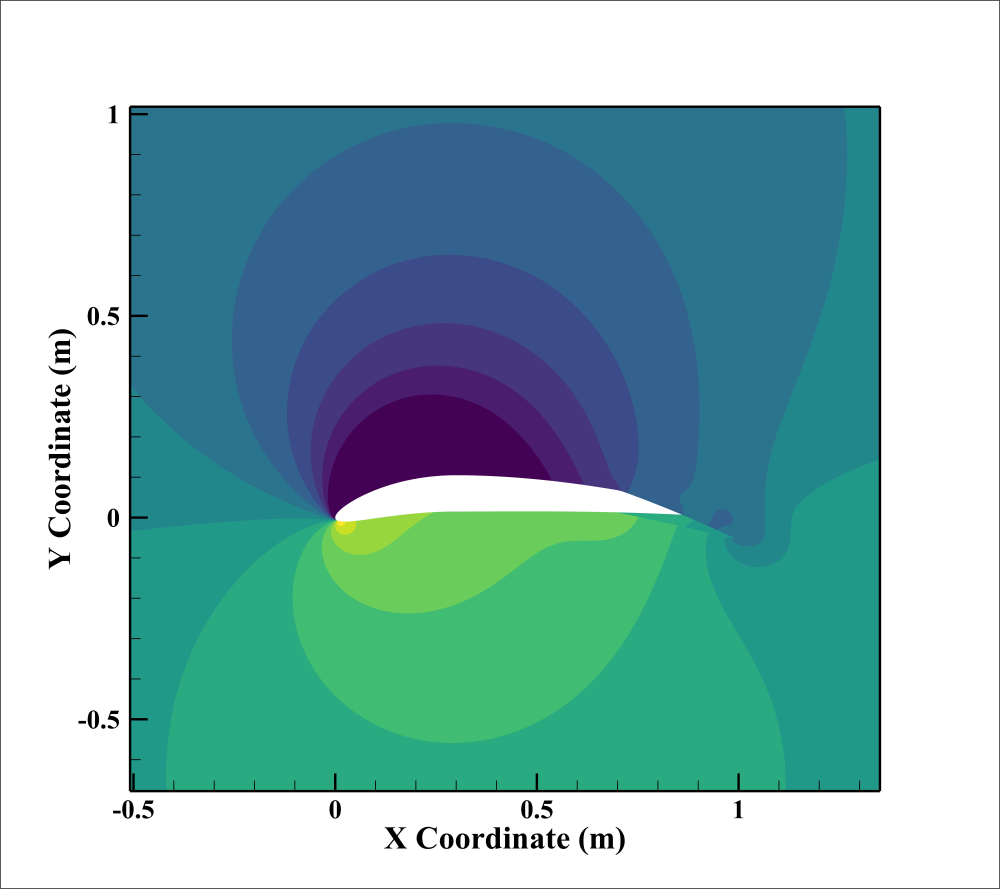}
            \caption{\(7^\circ\)}
        \end{subfigure}
        \hfill
        \begin{subfigure}[b]{0.31\textwidth}
            \includegraphics[width=\textwidth]{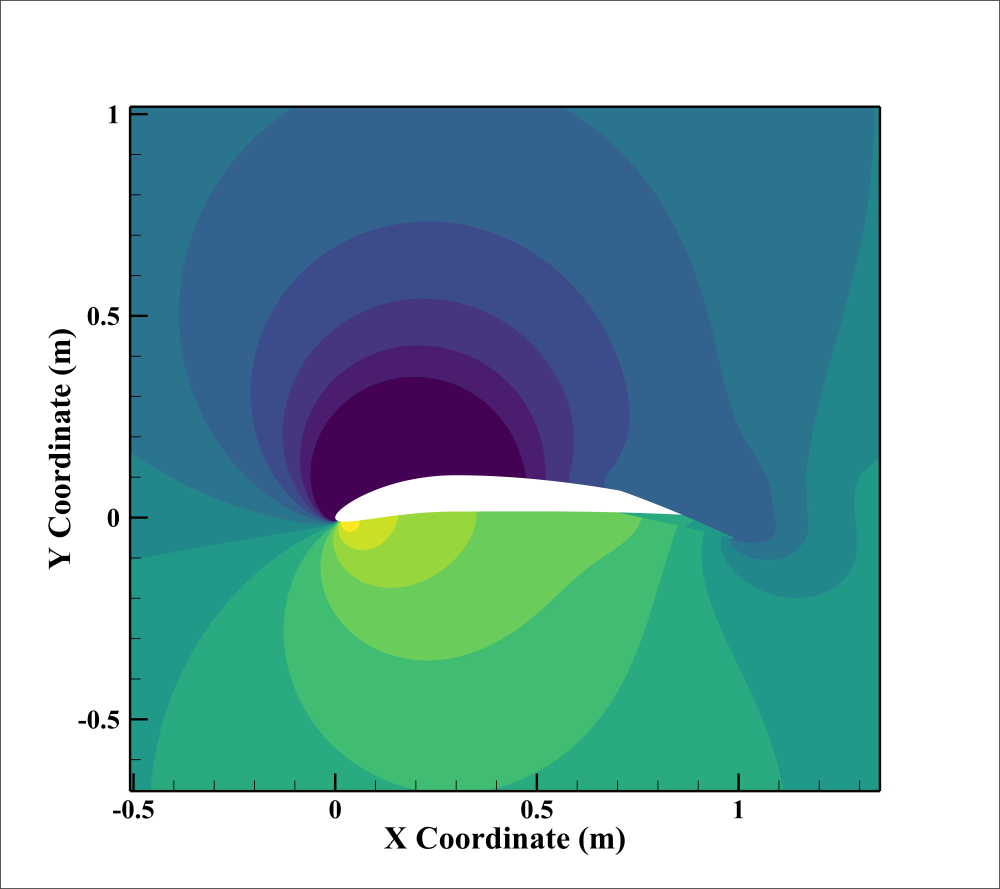}
            \caption{\(12^\circ\)}
        \end{subfigure}

        \vspace{1em}
        
        \begin{subfigure}[b]{0.31\textwidth}
            \includegraphics[width=\textwidth]{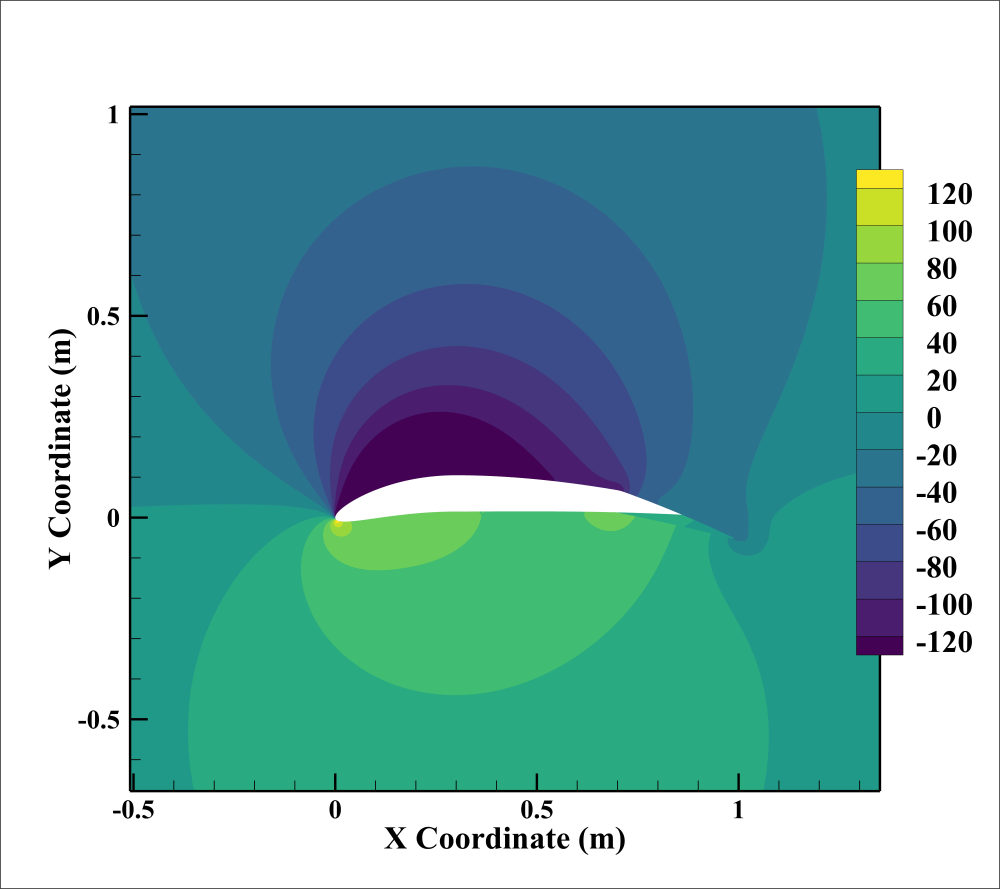}
            \caption{\(4^\circ\)}
        \end{subfigure}
        \hfill
        \begin{subfigure}[b]{0.31\textwidth}
            \includegraphics[width=\textwidth]{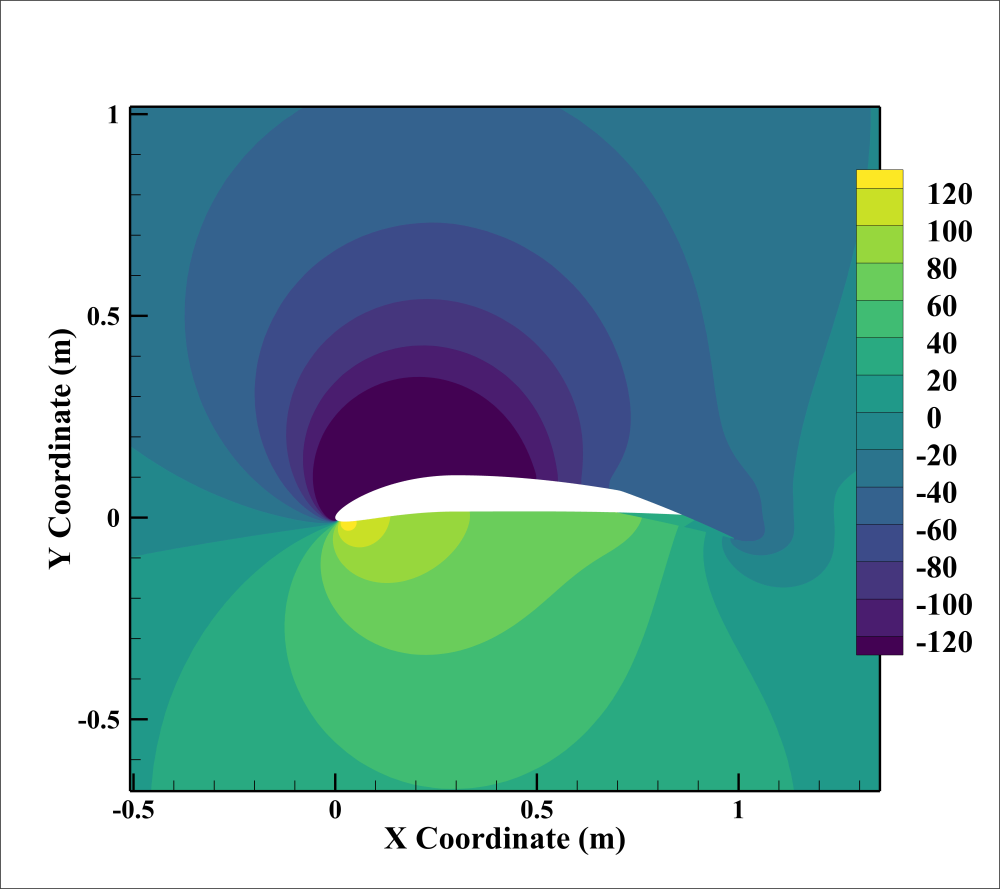}
            \caption{\(11^\circ\)}
        \end{subfigure}
        \hfill
        \begin{subfigure}[b]{0.31\textwidth}
            \includegraphics[width=\textwidth]{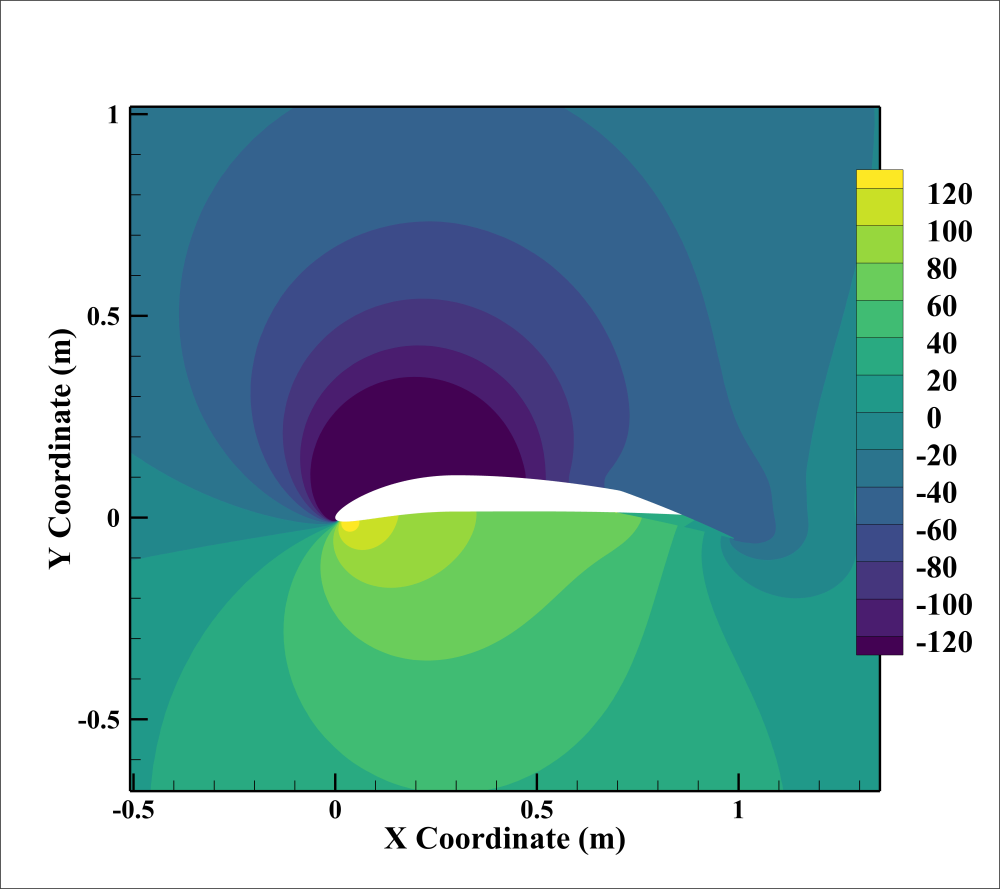}
            \caption{\(13^\circ\)}
        \end{subfigure}
    \end{subfigure}
\caption{\small Modified NACA6309 with \(10^\circ\) Down-Flap Pressure Contour at \(0^\circ\), \(4^\circ\), \(7^\circ\), \(12^\circ\), \(13^\circ\) angle of attack.}
    \label{fig:twosets}
\end{figure}

We can see from  section 4.1, that lift increases with a higher angle of attack and around \(13^\circ\) to \(14^\circ\), most of the stall occurs. This is synonymous with Fig. 8, where the static pressure contour is developed for NACA6309. With increasing angle of attack, the wholistic pressure distribution under the airfoil increased and overall, the pressure increased around the body. At, \(13^\circ\), the stall occurs and the phenomenon is clear from the static pressure contour. At the lower part of the airfoil, the pressure stayed within a magnitude of the maximum range, nevertheless in the upper portion the pressure distribution is way higher than other angles of attack, and the region is colored orange. After stalling the drag force drastically increased which is clear from Fig. 8. In Fig. 9 the modified NACA 6309 is given as having \(1^\circ\) up-flap. From the pressure contour, it is apparent that the lift performance is poorer than baseline NACA6309, as the developed pressure difference under and above the airfoil is relatively lower. But as the angle of attack increases, the pressure development increases, and at \(13^\circ\) angle of attack stall occurs. Nonetheless, before and after the stall, the leading edge and trailing edge pressure difference remained relatively low. Whereas, in other airfoils with increasing angle of attack, this particular pressure difference increases, increasing the overall drag and resulting reduction in power output.

But in the case of \(1^\circ\) up-flap, the drag force stayed sufficiently low; with higher lift generation, resulting in the optimal airfoil shape. In Fig. 10, the modified NACA6309 flap possessing \(10^\circ\) down-flap is presented. In this case, at 0° angle of attack the lower part of the airfoil had higher pressure than the upper skin, making the airfoil obtain higher lift force. But as the angle of attack increases because of the shape of the trailing edge being highly flapped, additional pressure generation occurred in that region, making the drag force higher on such a low angle of attack; suffering power output. As a result, with increasing angle of attack, the drag becomes extremely high. Hence, the overall lift-to-drag ratio reduces making the airfoil the least performing of them all. 
\begin{figure}[htbp]
    \centering
    \begin{subfigure}[h]{0.31\textwidth}
        \centering
      \includegraphics[width=\textwidth]{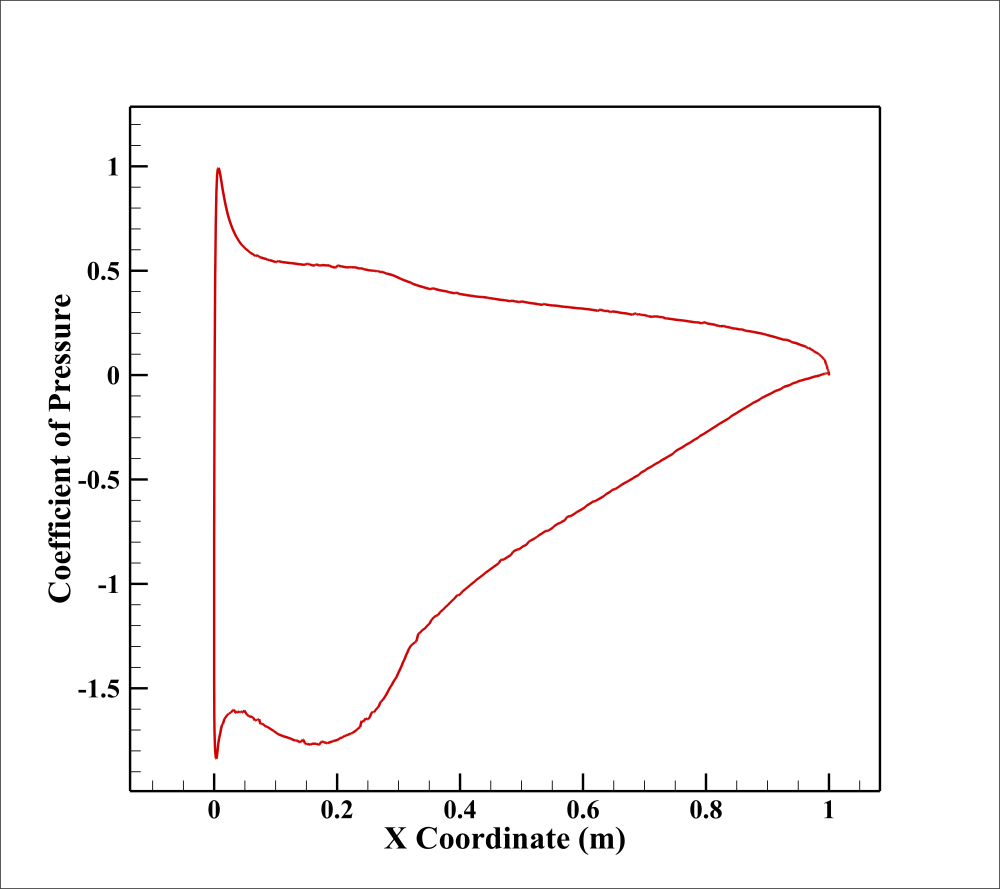}
        \caption{Base}
        \label{}
    \end{subfigure}
    \hfill
    \begin{subfigure}[h]{0.31\textwidth}
        \centering
        \includegraphics[width=\textwidth]{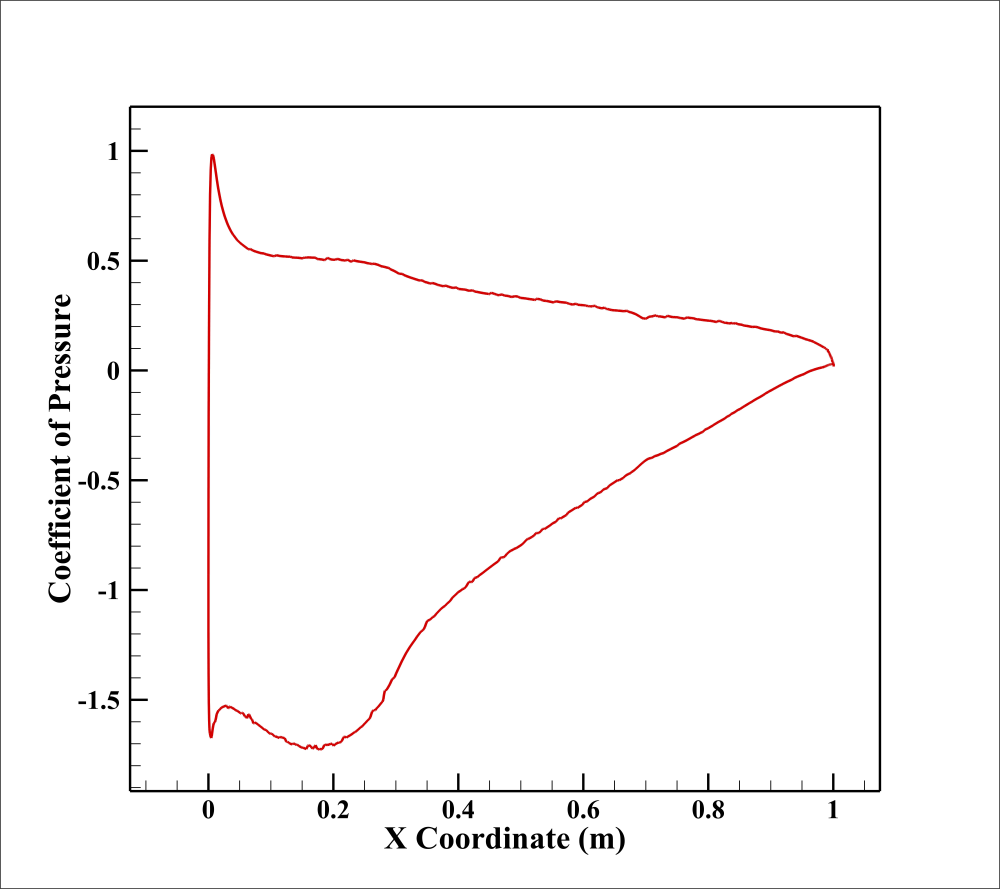}
        \caption{ \(1^\circ\) UF}
        \label{fig:subfig2}
  \end{subfigure}
  \hfill
    \begin{subfigure}[h]{0.31\textwidth}
        \centering
        \includegraphics[width=\textwidth]{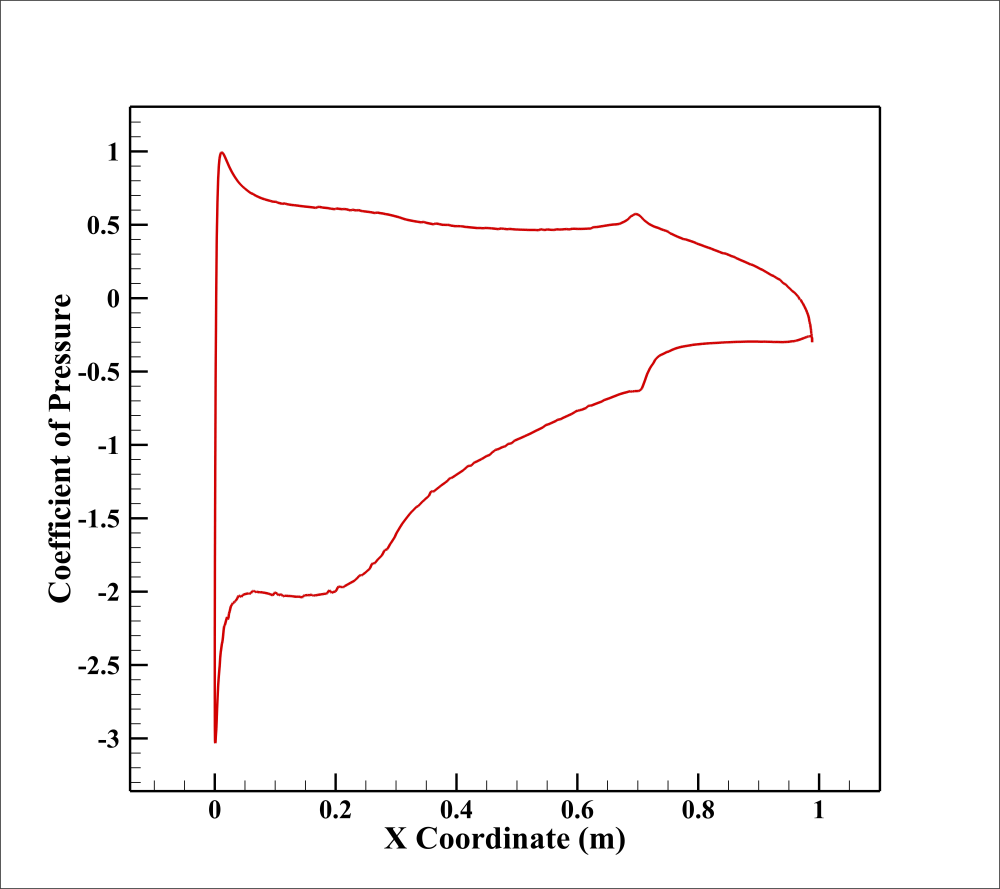}
        \caption{\(1^\circ\) DF}
        \label{fig:subfig2}
  \end{subfigure}
  \hfill
  \vspace{3em}
    \begin{subfigure}[h]{0.31\textwidth}
        \centering
        \includegraphics[width=\textwidth]{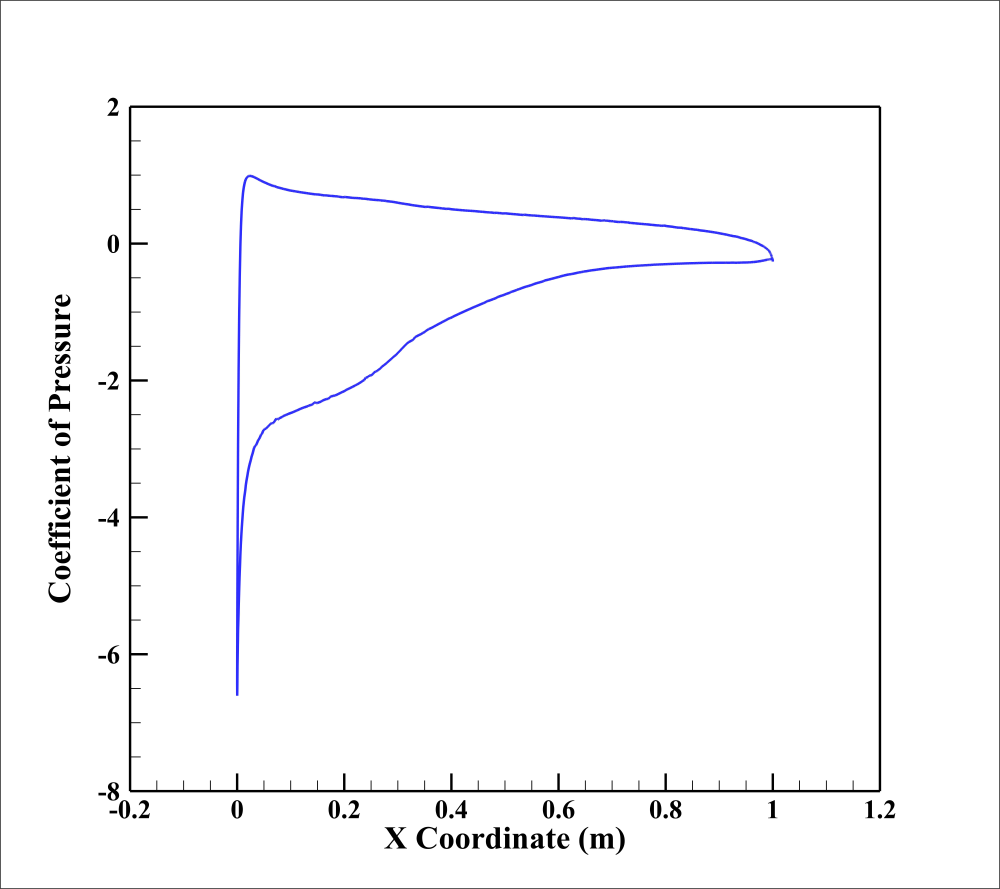}
        \caption{Base}
        \label{fig:subfig2}
  \end{subfigure}
  \hfill
    \begin{subfigure}[h]{0.31\textwidth}
        \centering
        \includegraphics[width=\textwidth]{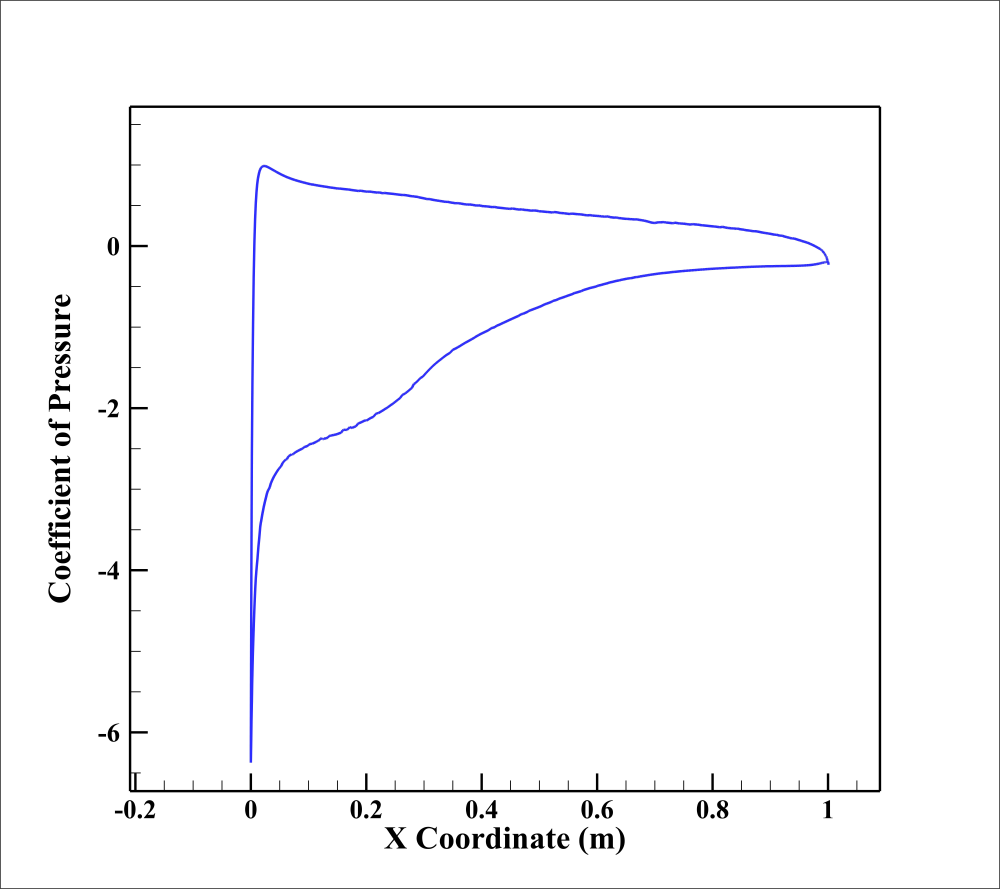}
        \caption{\(1^\circ\) UF}
        \label{fig:subfig2}
  \end{subfigure}
  \hfill
    \begin{subfigure}[h]{0.31\textwidth}
        \centering
        \includegraphics[width=\textwidth]{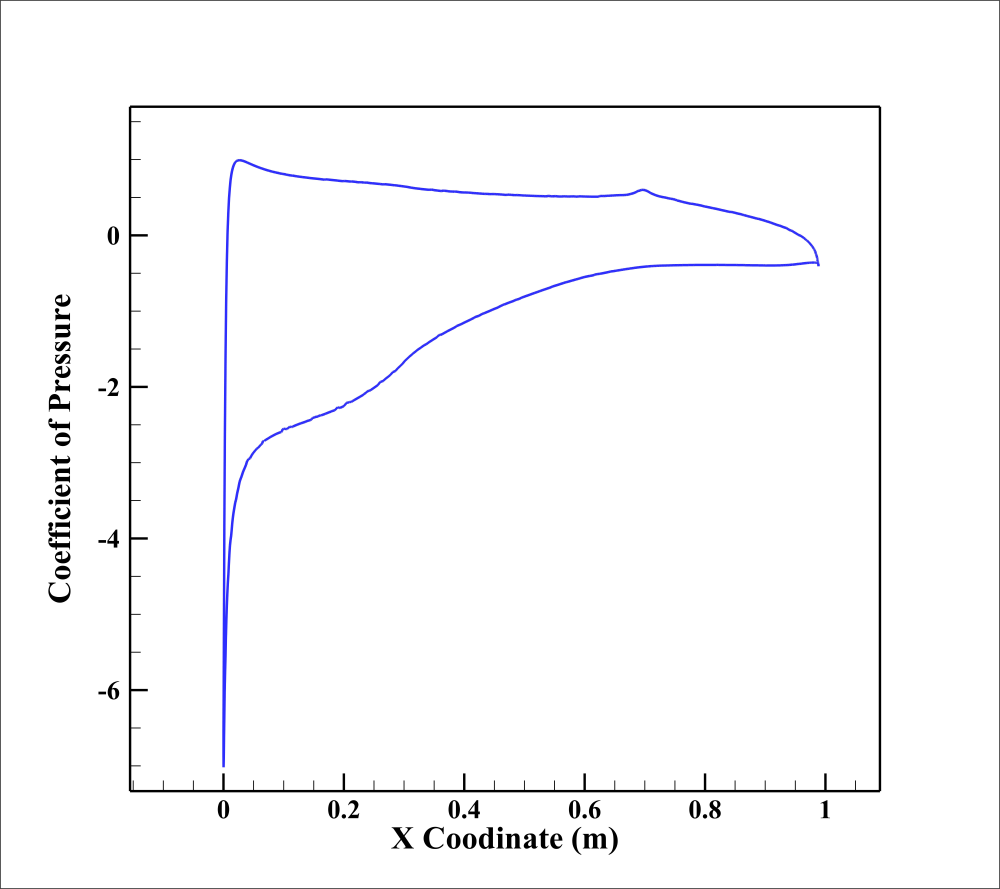}
        \caption{\(10^\circ\) DF}
        \label{fig:subfig2}
  \end{subfigure}
    \caption{\small Coefficient of pressure of NACA6309 baseline, 1UF, 10DF at \(7^\circ\) and \(13^\circ\) angle of attack.}
    \label{fig:sidebyside}
\end{figure}
\subsection{Investigation of Coefficient of Pressure}
\setlength{\parindent}{0.2in}
\hspace{0.2in}
The Pressure coefficient is a dimensionless number that creates the relationship between relative pressure and static pressure around a body subjected to fluid flow. In Fig. 11, the pressure coefficient of NACA 6309 baseline, one-degree up-flap, and ten-degree down-flap is given at \(7^\circ\) and \(13^\circ\) angle of attack, because of at \(7^\circ\) angle of attack the lift-drag performance is quite stable and at \(13^\circ\) angle of attack stall occurs. In the baseline airfoil, at  \(7^\circ\) angle of attack the relative pressure difference is higher creating sufficient lift, whereas at \(13^\circ\) angle of attack the pressure difference gets squeezed at the trailing edge with respect to static pressure. In the one-degree up-flap configuration, at \(7^\circ\)  angle of attack, having a slight up flap at the trailing edge, the pressure difference exhibits a lower magnitude than baseline NACA6309. Moreover, at the leading edge, the coefficient of pressure dropped lower than NACA6309 baseline. On the other hand, in a ten-degree down-flap configuration, at \(7^\circ\)  angle of attack, the pressure difference exhibits a higher magnitude among the three. At \(13^\circ\) angle of attack, the stall occurs in the case of the one-degree up-flap, therefore the pressure difference with respect to static pressure is seen to be compromised. In the case of a ten-degree down-flap similar phenomenon is seen after stalling.  

\subsection{Investigation of Velocity}
\setlength{\parindent}{0.2in}
\hspace{0.2in}
To properly investigate the airfoil performance, we need to dive into the velocity distribution of the flow around the airfoil. In Fig. 12, 13, and 14 the velocity distribution around the airfoil for baseline and two modified airfoils: best which is the one-degree up-flap, and worst which is the ten-degree down-flap. From Fig. 12 it can be seen that the flow separation region at the trailing edge increases as the angle of attack increases. At the initial angle of attack trials, the velocity was sufficient. But as the angle of attack increases, the flow velocity continuously decreases and in the upper part of the airfoil, the velocity results were comparatively better which results in lower performance overall impeding power generation. Furthermore, from Fig. 13, it can be stated that at 0° angle of attack, two small vortices are observed at the leading edge of the airfoil, one is the comparatively lower velocity region (the green one), and the other is the higher velocity region (orange one). Also, at the lower angle of attacks, the flow separation region is relatively smaller than the baseline airfoil. But like the baseline, with the increasing angle of attack, the flow separation region increased. Nevertheless, the velocity flow separation occurs properly at stall angle, until then the flow held streamlined properties. This is because the upward flap possessed velocity-recovering characteristics. On the other hand, from Fig. 14, it is clear that at the trailing edge, the velocity separation occurred in an early stage. With increasing angle of attack, the flow separation region dramatically increases, surfeiting the other airfoil performance. In addition, lower velocity regions can be visible in the lower part of the airfoil. This is mainly because of the downward arrangement of the flap. The downward angle creates additional flow deterioration of the velocity performance.

\begin{figure}[htbp]
    \centering
    \begin{subfigure}[h]{0.31\textwidth}
        \centering
      \includegraphics[width=\textwidth]{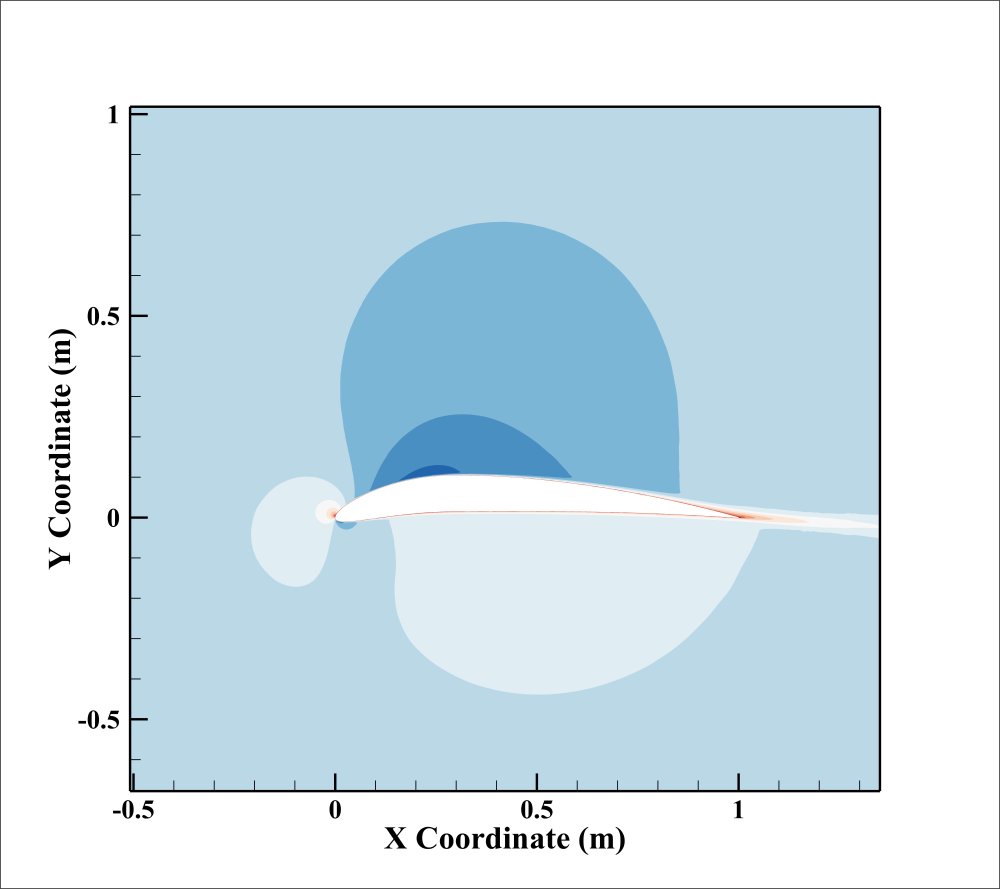}
        \caption{\(0^\circ\)}
        \label{}
    \end{subfigure}
    \hfill
    \begin{subfigure}[h]{0.31\textwidth}
        \centering
        \includegraphics[width=\textwidth]{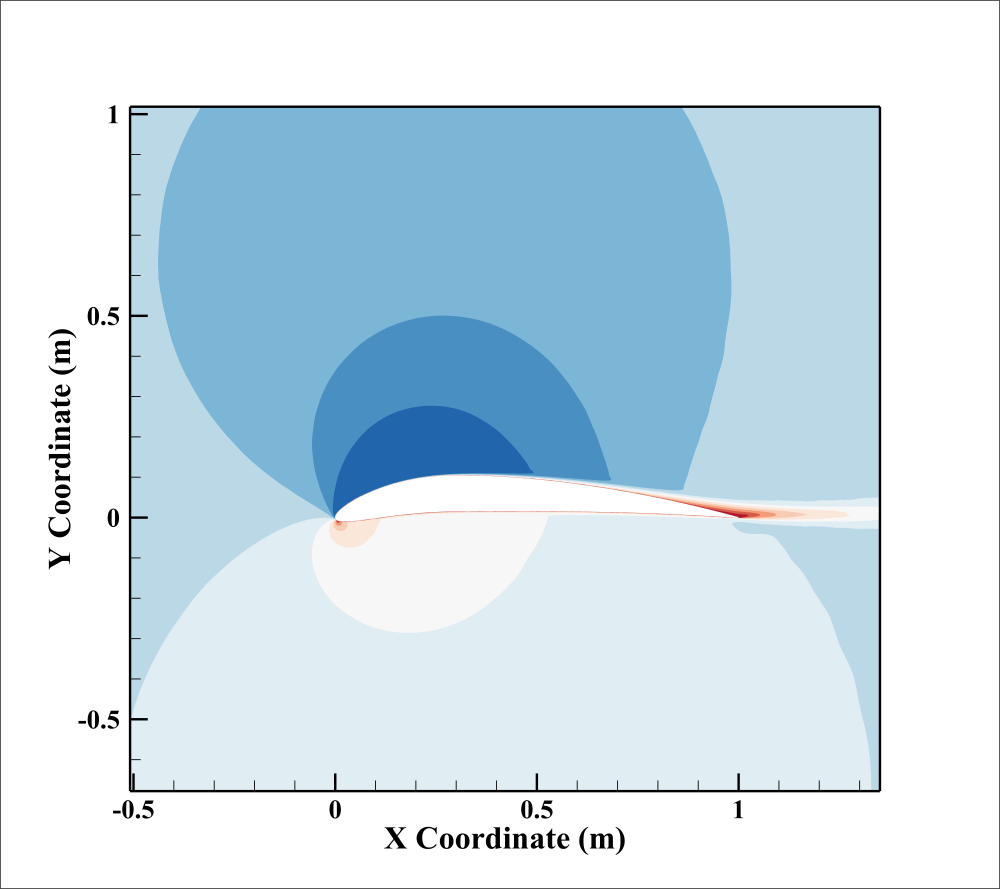}
        \caption{\(7^\circ\)}
        \label{fig:subfig2}
  \end{subfigure}
  \hfill
    \begin{subfigure}[h]{0.31\textwidth}
        \centering
        \includegraphics[width=\textwidth]{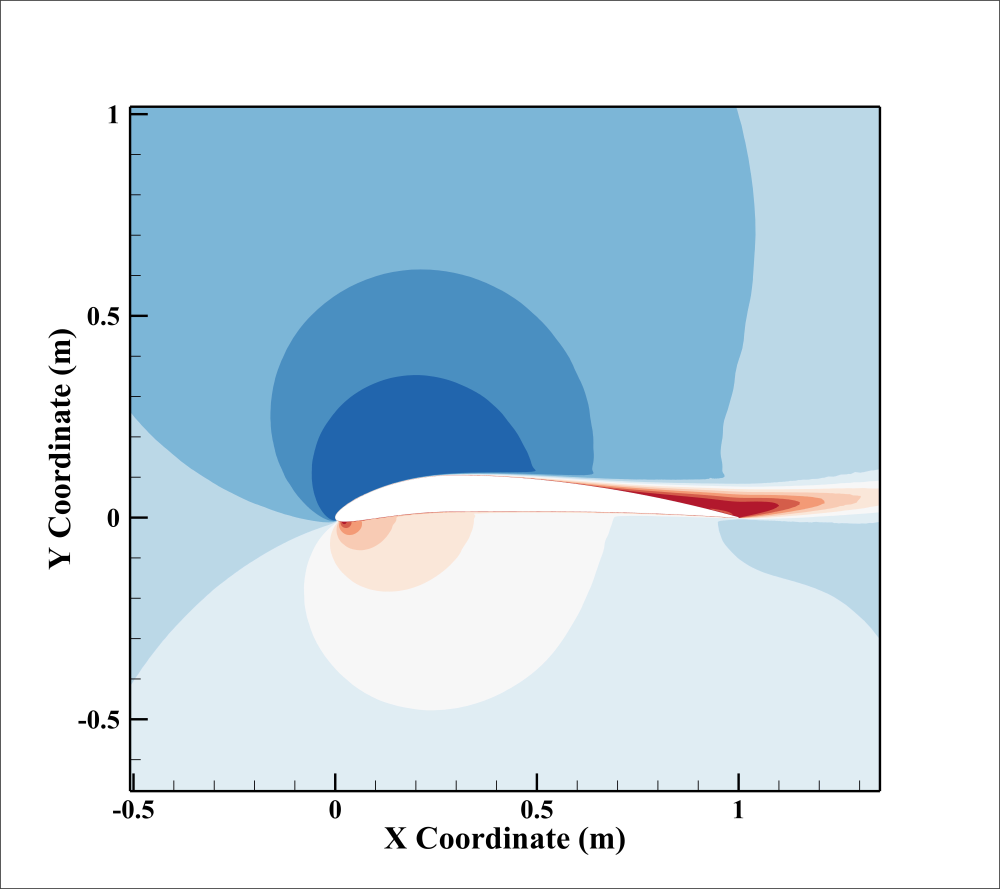}
        \caption{\(12^\circ\)}
        \label{fig:subfig2}
  \end{subfigure}
  \hfill
  \vspace{3em}
    \begin{subfigure}[h]{0.31\textwidth}
        \centering
        \includegraphics[width=\textwidth]{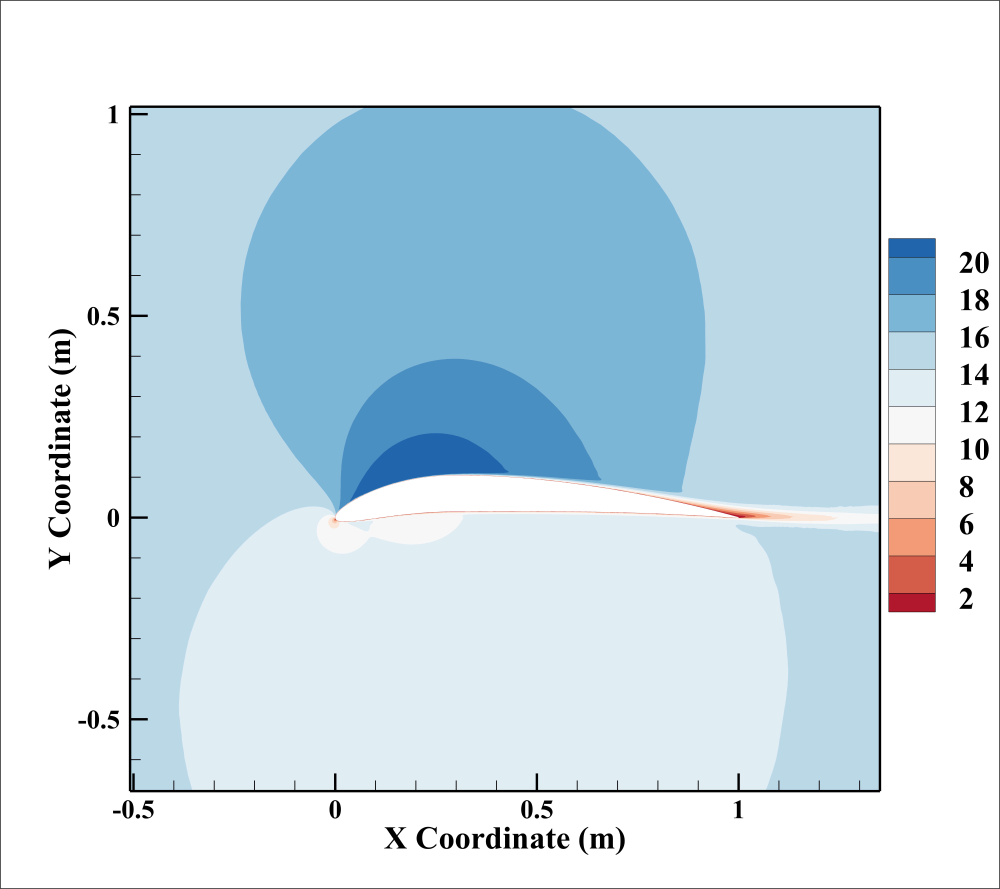}
        \caption{\(4^\circ\)}
        \label{fig:subfig2}
  \end{subfigure}
  \hfill
    \begin{subfigure}[h]{0.31\textwidth}
        \centering
        \includegraphics[width=\textwidth]{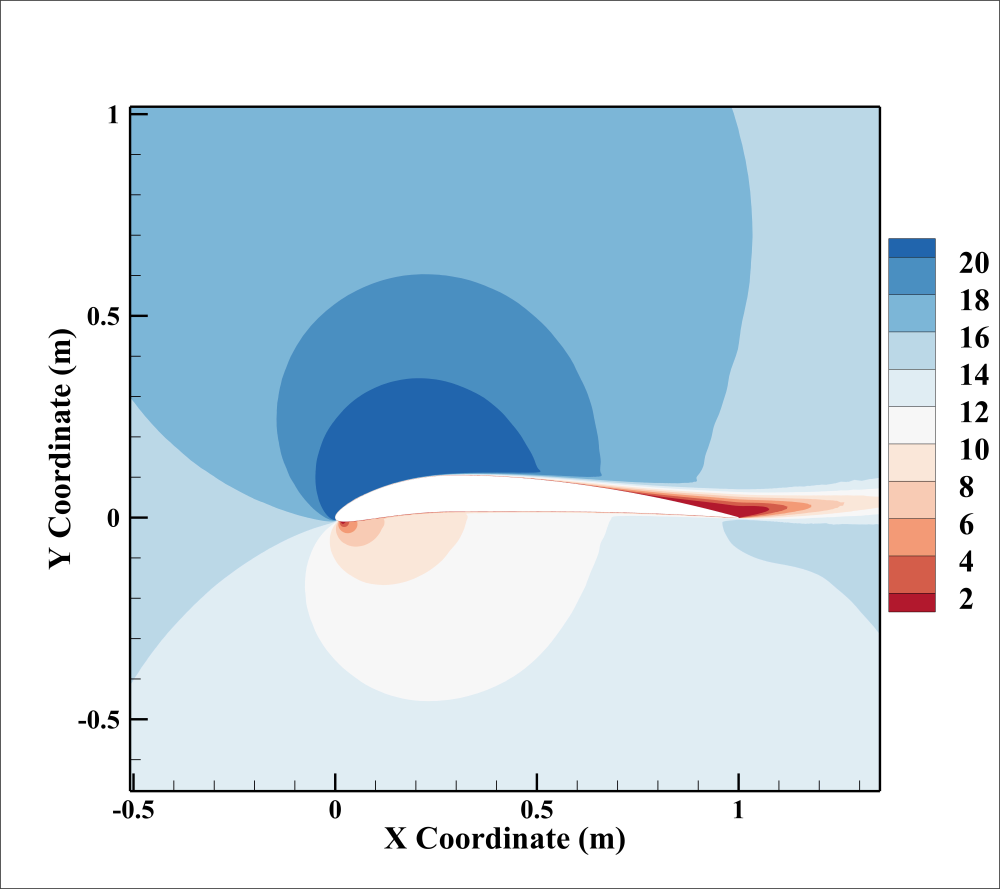}
        \caption{\(11^\circ\)}
        \label{fig:subfig2}
  \end{subfigure}
  \hfill
    \begin{subfigure}[h]{0.31\textwidth}
        \centering
        \includegraphics[width=\textwidth]{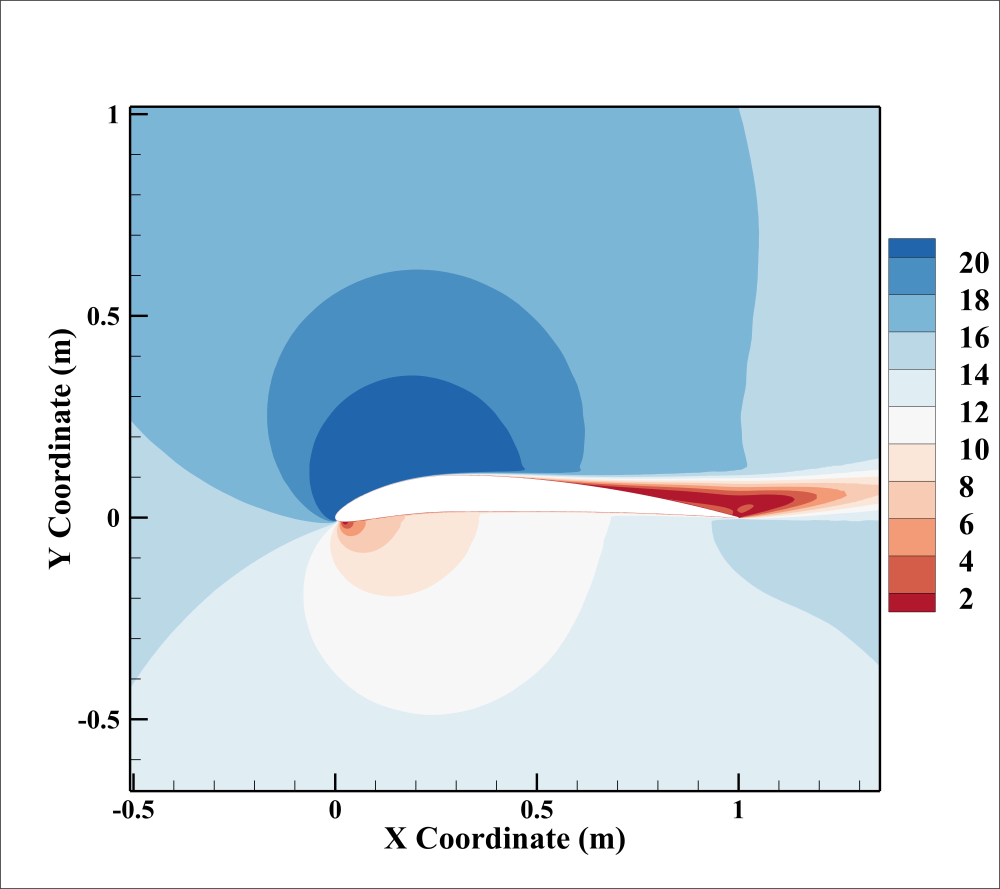}
        \caption{\(13^\circ\)}
        \label{fig:subfig2}
  \end{subfigure}
    \caption{\small NACA6309 with Velocity Contour at \(0^\circ\), \(4^\circ\), \(7^\circ\), \(12^\circ\), \(13^\circ\) angle of attack.}
    \label{fig:sidebyside}
\end{figure}

 \begin{figure}[htbp]
    \centering
    \begin{subfigure}[b]{\textwidth}
        \centering
        \begin{subfigure}[b]{0.31\textwidth}
            \includegraphics[width=\textwidth]{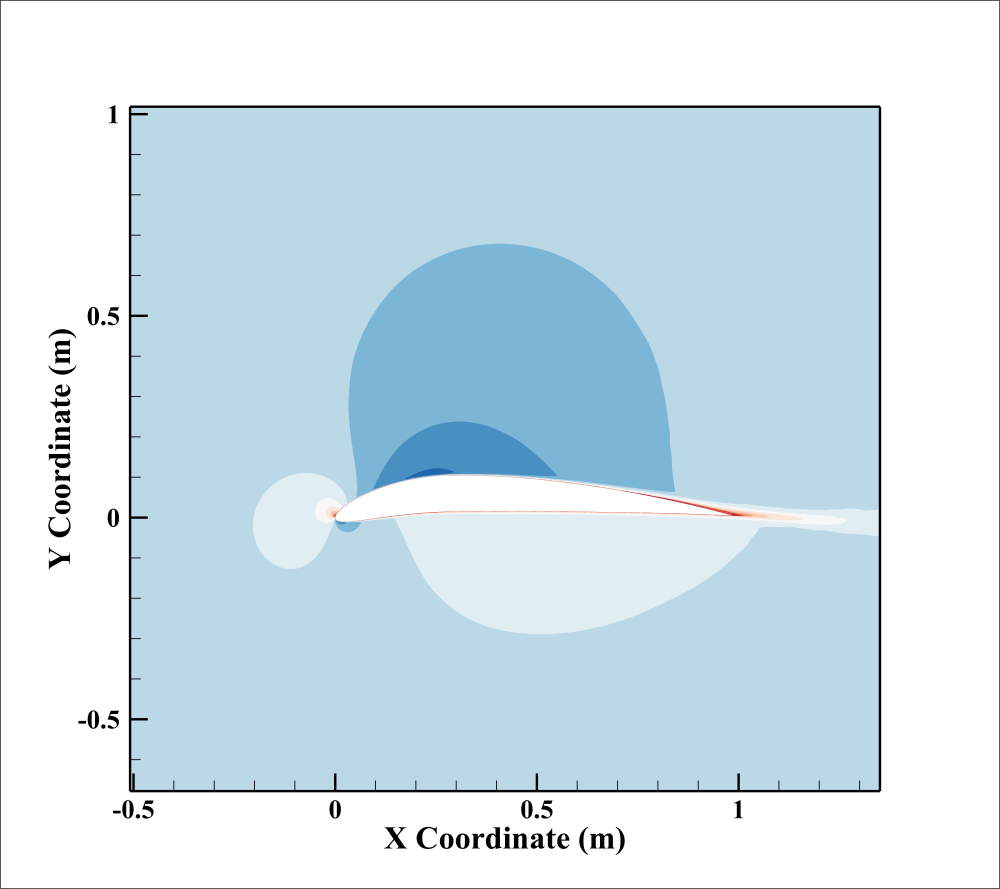}
            \caption{\(0^\circ\)}
        \end{subfigure}
        \hfill
        \begin{subfigure}[b]{0.31\textwidth}
            \includegraphics[width=\textwidth]{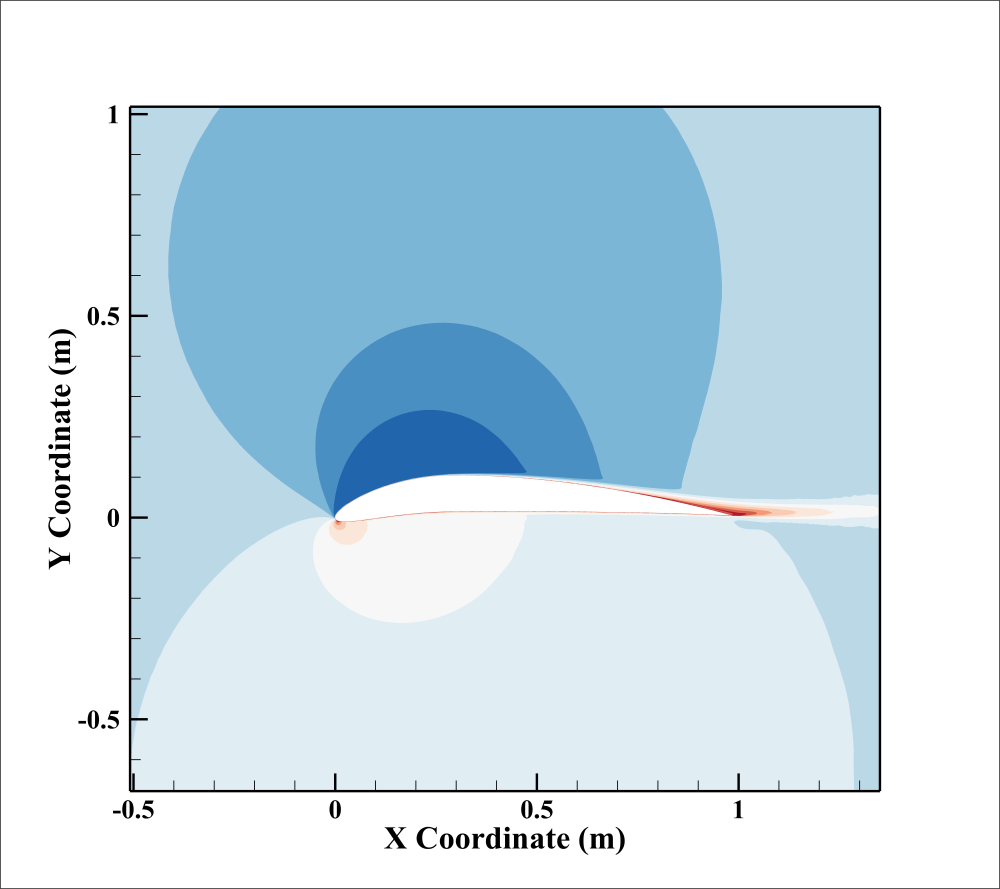}
            \caption{\(7^\circ\)}
        \end{subfigure}
        \hfill
        \begin{subfigure}[b]{0.31\textwidth}
            \includegraphics[width=\textwidth]{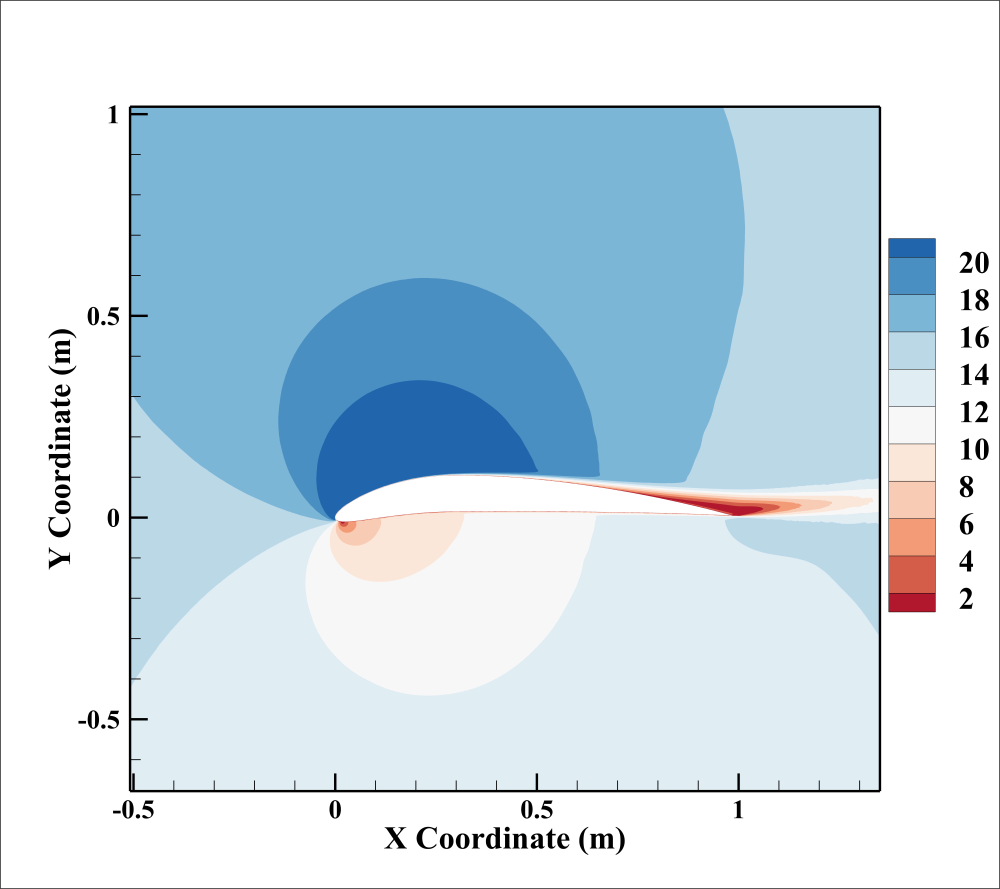}
            \caption{\(12^\circ\)}
        \end{subfigure}
        
        \vspace{1em}
        
        \begin{subfigure}[b]{0.31\textwidth}
            \includegraphics[width=\textwidth]{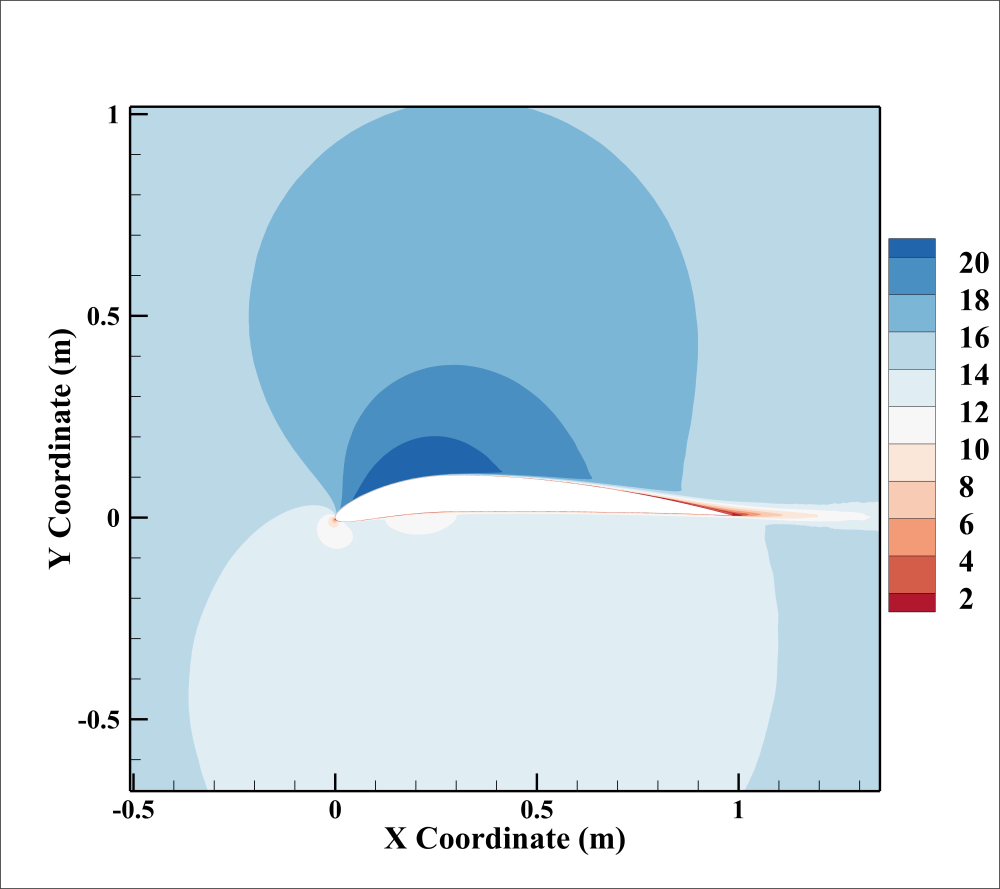}
            \caption{\(4^\circ\)}
        \end{subfigure}
        \hfill
        \begin{subfigure}[b]{0.31\textwidth}
            \includegraphics[width=\textwidth]{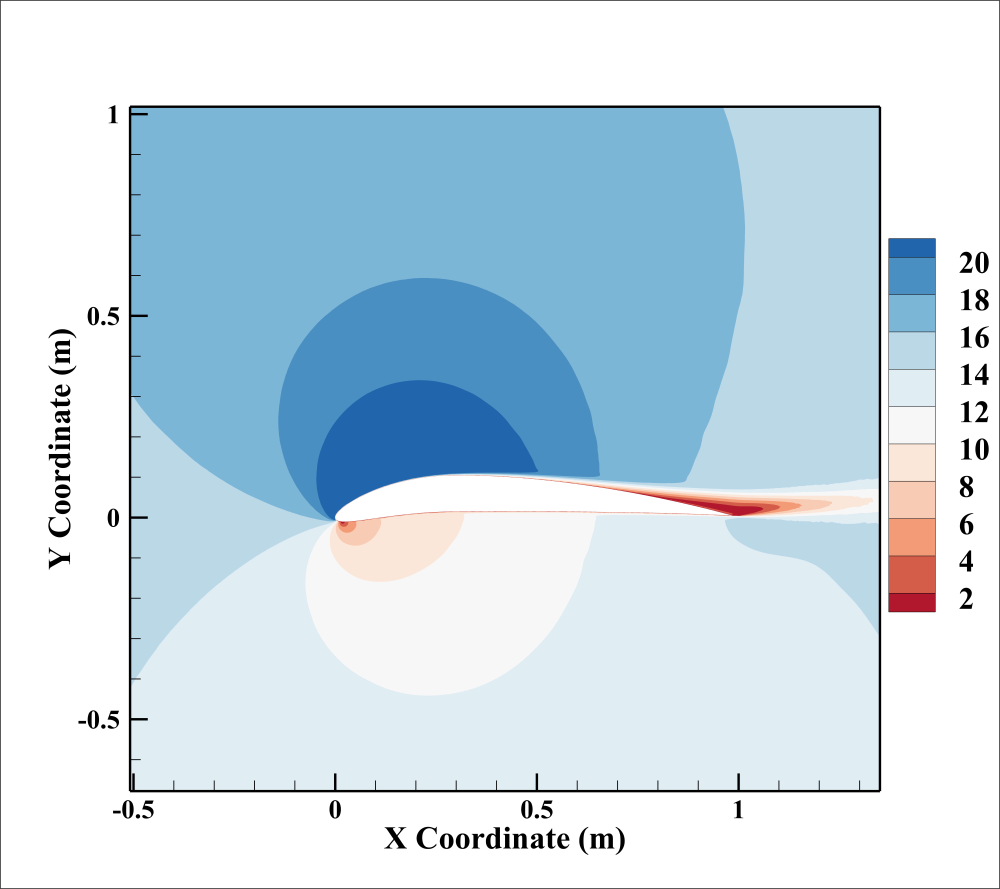}
            \caption{\(11^\circ\)}
        \end{subfigure}
        \hfill
        \begin{subfigure}[b]{0.31\textwidth}
            \includegraphics[width=\textwidth]{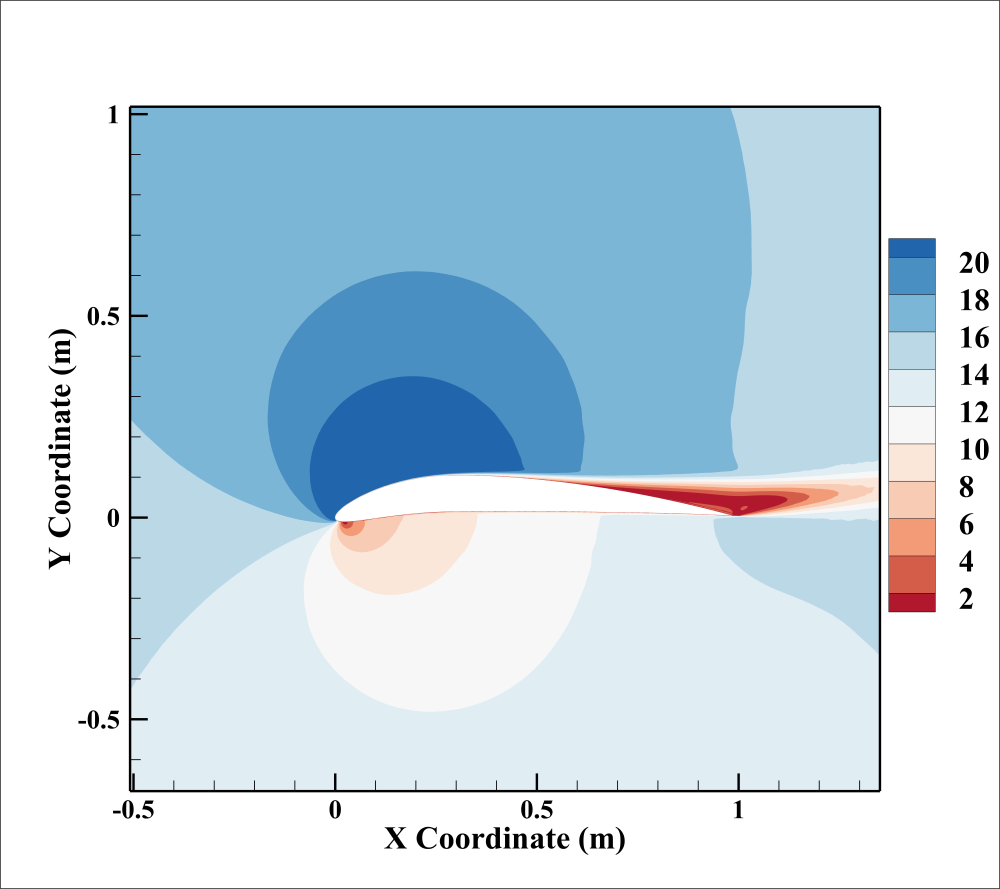}
            \caption{\(13^\circ\)}
        \end{subfigure}
    \end{subfigure}
    \caption{\small Modified NACA6309 with 1 Degree Up-flap Velocity Contour at \(0^\circ\), \(4^\circ\), \(7^\circ\), \(12^\circ\), \(13^\circ\) angle of attack.}
    \vspace{2em} 

    \setcounter{subfigure}{0}

    \begin{subfigure}[b]{\textwidth}
        \centering
        \begin{subfigure}[b]{0.31\textwidth}
            \includegraphics[width=\textwidth]{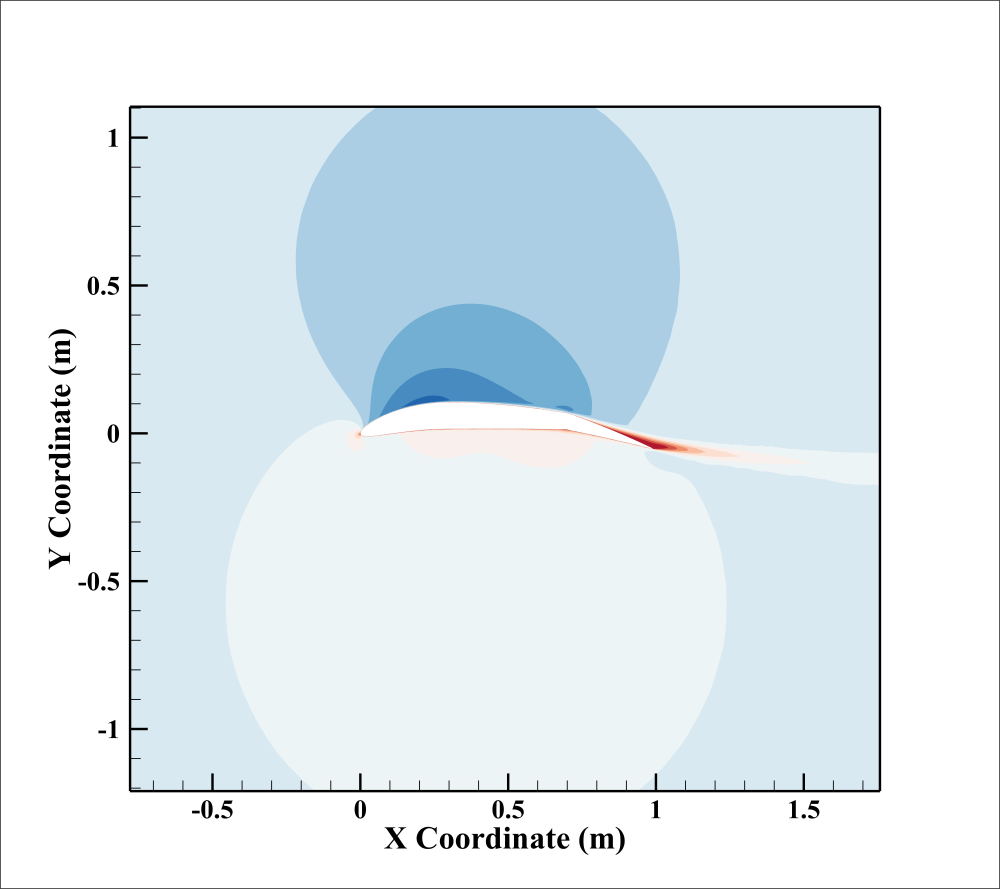}
            \caption{\(0^\circ\)}
        \end{subfigure}
        \hfill
        \begin{subfigure}[b]{0.31\textwidth}
            \includegraphics[width=\textwidth]{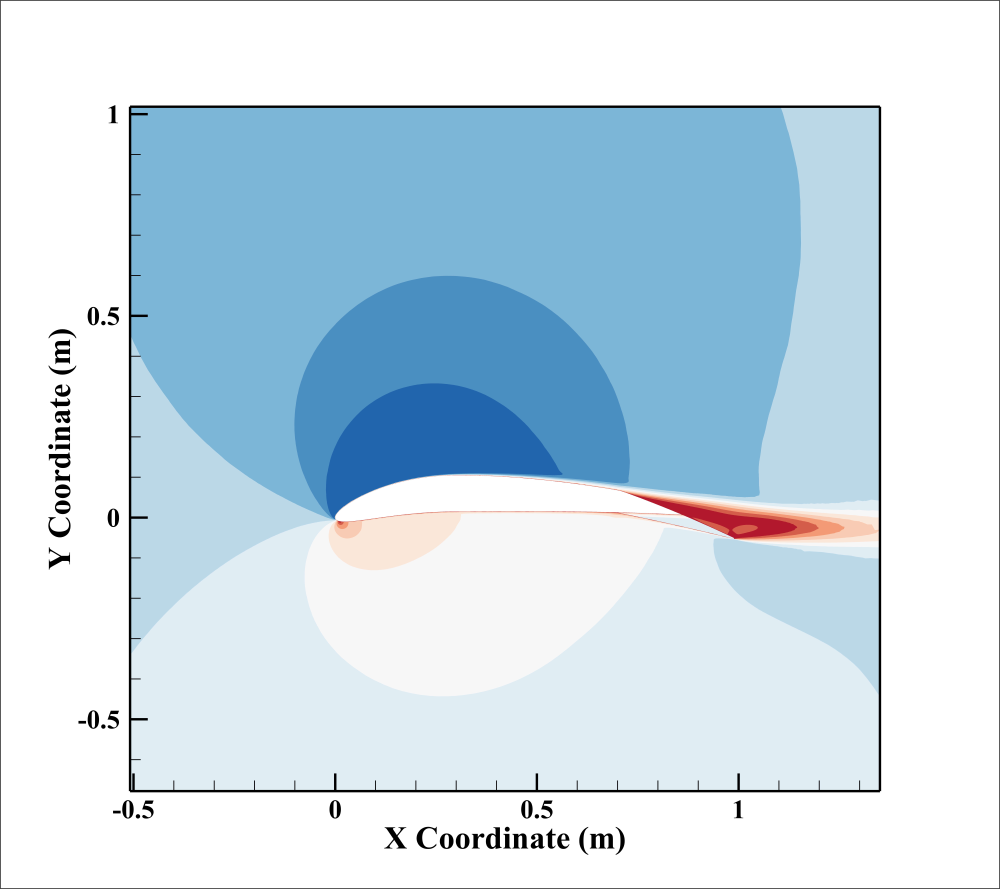}
            \caption{\(7^\circ\)}
        \end{subfigure}
        \hfill
        \begin{subfigure}[b]{0.31\textwidth}
            \includegraphics[width=\textwidth]{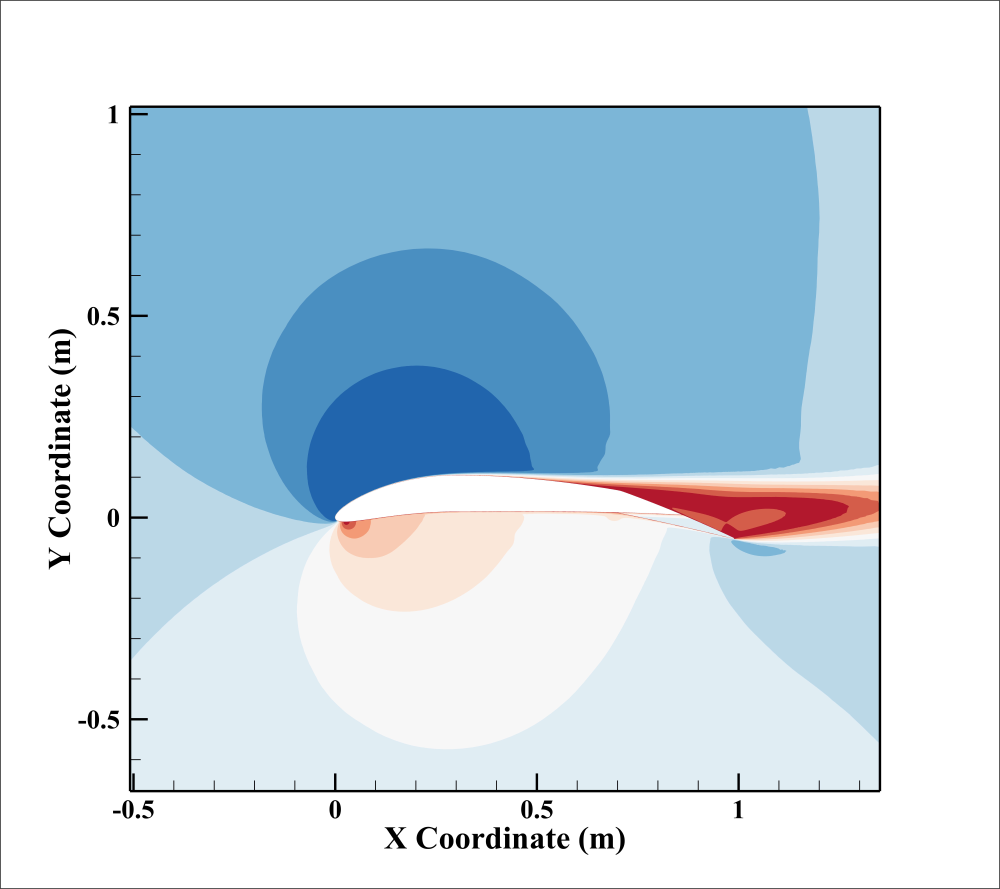}
            \caption{\(12^\circ\)}
        \end{subfigure}

        \vspace{1em}
        
        \begin{subfigure}[b]{0.31\textwidth}
            \includegraphics[width=\textwidth]{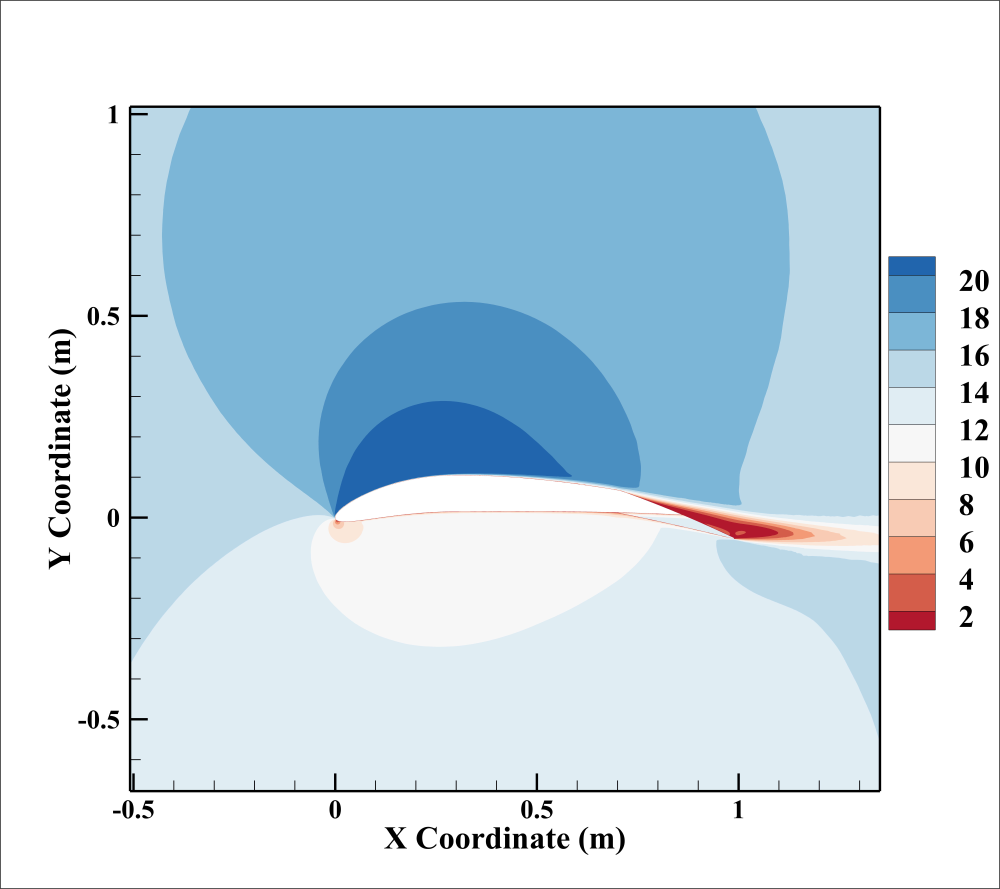}
            \caption{\(4^\circ\)}
        \end{subfigure}
        \hfill
        \begin{subfigure}[b]{0.31\textwidth}
            \includegraphics[width=\textwidth]{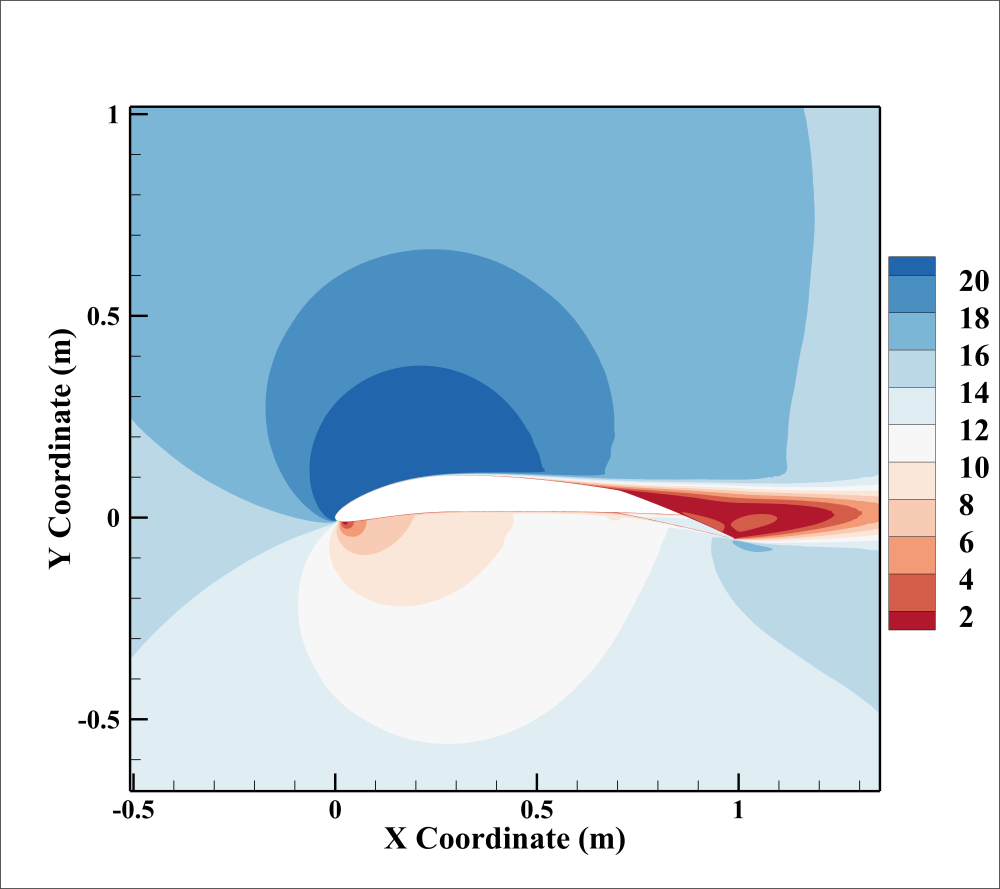}
            \caption{\(11^\circ\)}
        \end{subfigure}
        \hfill
        \begin{subfigure}[b]{0.31\textwidth}
            \includegraphics[width=\textwidth]{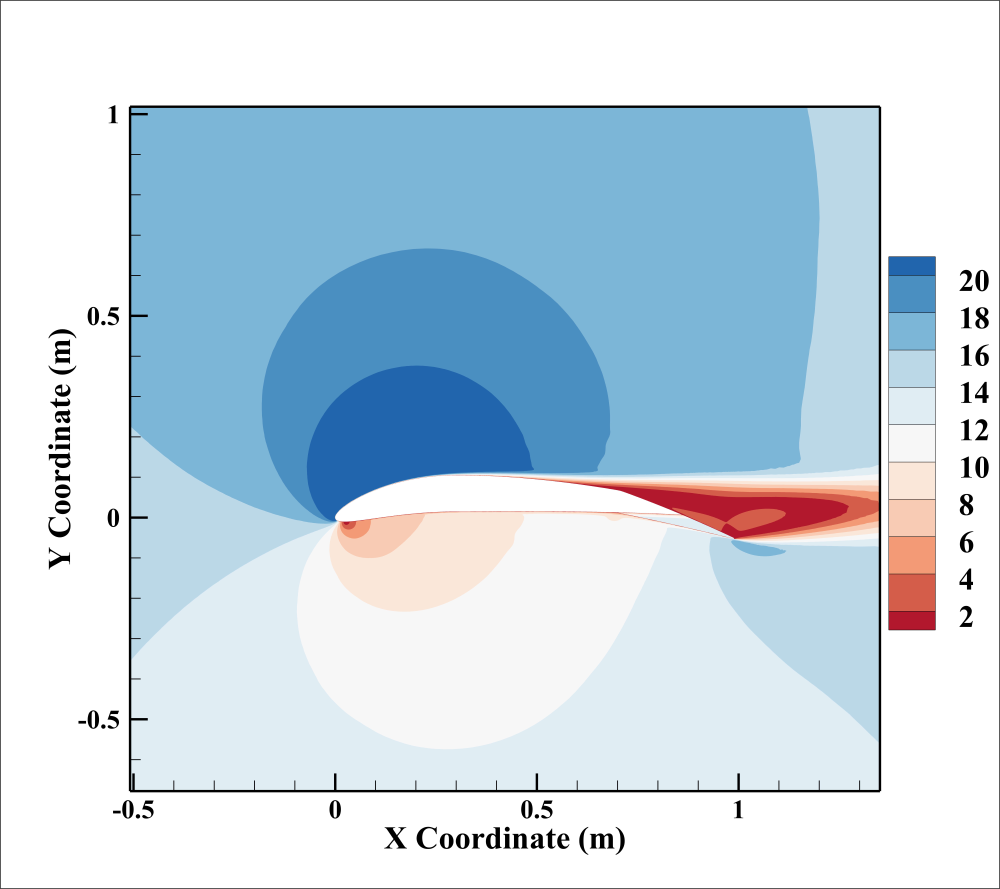}
            \caption{\(13^\circ\)}
        \end{subfigure} 
    \end{subfigure}
    \caption{\small Modified NACA6309 with 10 Degree Down-flap Velocity Contour at \(0^\circ\), \(4^\circ\), \(7^\circ\), \(12^\circ\), \(13^\circ\) angle of attack.}
    \label{fig:twosets}
\end{figure}
\newpage
\section{Conclusion}
\setlength{\parindent}{0.2in}
\hspace{0.2in}
The use of flap acting as a passive flow-controlling device has been demonstrated by an in-depth numerical study. The influence of flap angles on the upward and downward sides was analyzed for NACA6309 airfoil. The turbulence model has been chosen as Spalart-Allmaras. Based on numerical data, it is evident that the turbulent model shows good accuracy. The influence of the flap was determined by tilting the flap in the upward and downward directions. In the upward direction, the flap angle was from \(1^\circ\) to \(5^\circ\) and in the downward direction, the flap angle was from \(1^\circ\) to \(10^\circ\). The analysis revealed that the presence of the flap leads to improvement of aerodynamic performance in the overall lift-to-drag ratio with the angle of attack maximizing turbine power output. Finally, to conclude the study:
\begin{itemize}
    \item An in-depth analysis of NACA6309 airfoil has been studied, which was not available in the literature until now. The lift, drag, lift to drag ratio with angle of attack have been studied which is vital for power generation of wind turbine blades having such airfoil. From the pressure and velocity contour, we got the overall aerodynamic performance. The lift and drag performance seem generally up to standard for proper wind turbine operation. 
    \item At  0° angle of attack, the lift performance was better in all cases of down-flap arrangement, in addition with increasing flap-angle the lift force increases, and at \(10^\circ\) down-flap arrangement it surpluses all. The up-flap arrangement initially performed poorer, but with increasing angle of attack, it catches up with most of the high lift generating airfoils. 
	
 \item With increasing down-flap angle the drag force increase drastically, resulting in powergeneration retardance. At \(10^\circ\) down-flap the drag force becomes maximum. Also, on the flip side, drag is incredibly lower at up-flap configurations. At \(1^\circ\) up-flap configuration, the drag force was the least of them all resulting in higher power generation. 
 \item As conflict of performance arises, the down-flap gave better lift results and the up-flap gave better drag results. It is quite apparent that, combining the two forces, from \(1^\circ\) up-flap configuration, the best performance was achieved. Though at 0° angle of attack the lift was relatively low, but with increasing angle of attack it catches up to better performance. Moreover, the \(1^\circ\) up-flap airfoil gave the best drag results, creating the epitome of power generation performance among all. 
 \item 	At NACA6309 airfoil, at \(7^\circ\) angle of attack, the relative pressure difference is quite higher creating sufficient lift, whereas at \(13^\circ\) angle of attack stall occurs, creating a squeeze in pressure difference at the trailing edge. 
 \item The more the flap angle increases downward the lower the performance. Slight upward configuration happens to be the most suitable candidate for higher power generation. 
\end{itemize}
	 
Finally, we can conclude that using the slight up-flap configuration in an airfoil is highly recommended for overall aerodynamic performance. This claim is not only coming from the lift-to-drag ratio but from the in-depth analysis of pressure and velocity distribution around the airfoil. 

\bibliographystyle{unsrtnat}
\bibliography{references}  






\end{document}